\documentclass[11pt,a4paper]{article}
\pdfoutput=1
\usepackage{jheppub}
\usepackage[utf8]{inputenc}
\usepackage{amsfonts}
\usepackage{amsmath}
\usepackage{color}
\usepackage[normalem]{ulem}
\usepackage{verbatim}
\usepackage{subcaption}
\usepackage{braket}

\preprint{UWTHPH-2020-87, MCnet-22-07}

\title{Colour Evolution and Infrared Physics}

\author[a,b,c]{Simon Pl\"atzer}
\affiliation[a]{Institute of Physics, NAWI Graz, University of Graz, Universit\"atsplatz 5, A-8010 Graz, Austria}
\affiliation[b]{Particle Physics, Faculty of Physics, University of Vienna, Boltzmanngasse 5, A-1090 Wien, Austria}
\affiliation[c]{Erwin Schr\"odinger Institute for Mathematics and Physics, University of Vienna, A-1090 Wien, Austria}
\emailAdd{simon.plaetzer@uni-graz.at}

\abstract{We give a complete account of how soft gluon, massless
    quark, evolution equations in colour space originate, from a
    factorization into a hard cross section density operator and a
    soft function encoding measurements and the projection on definite
    colours. We detail this formalism up to the two loop level and we
    demonstrate how the evolution kernels relate to infrared
    subtractions, and how the resolution of infrared singular regions
    conspires with the structure of observables the algorithm should
    be able to predict. The latter allows us to address evolution in
    different kinematic variables, including energy ordering and
    angular cutoffs in non-global observables. The soft factor and its
    evolution resembles a hadronization model including effects such
    as colour reconnection, and could give insight into the structure
    of power corrections in observables which require soft gluon
    evolution.}

\begin{document}

\maketitle

\section{Introduction}

The reliable prediction of realistic final states produced in high
energy collisions is at the heart of both, event generator simulation
as well as, in a less versatile but typically more precise fashion,
analytic resummation. Recent work has been highlighting the basic fact
that both approaches need to rely on analyzing amplitudes with many
external particles in order to fully account how the final state
builds up in detail. Cross sections are then obtained by squaring such
an amplitude and in the presence of a (certain class of) measurements
leading contributions can then be accounted for. A simplification is
possible by using QCD coherence for the resummation of global
observables, or dipole algorithms suited for non-global observables in
the large-$N$ limit, though in general more complicated evolution
equations remain to be addressed.

Parton branching at the amplitude level
\cite{Nagy:2014mqa,Nagy:2017ggp,Forshaw:2019ver} has become an
important theoretical framework to construct and analyze parton shower
algorithms. These approaches generalize the soft gluon evolution
algorithm discussed in \cite{Martinez:2018ffw}, which has recently
been used to resum non-global logarithms beyond the leading-$N$ limit
\cite{Platzer:2013fha,DeAngelis:2020rvq}.  Understanding the
structures in the soft limit unveils most of the complexity we need to
account for in improved algorithms since collinear physics is
typically colour diagonal and only soft physics can guide us to what
structures new parton showers need to build on
\cite{Holguin:2020joq,Holguin:2020oui}. In particular, we have
recently been pointing out that there exist colour-diagonal, though
formally $1/N$ suppressed contributions in the two-loop and
one-loop/one-emission contributions \cite{Platzer:2020lbr}. In the
same work, we have also laid the ground to analyse virtual corrections
differential in phase-space type integrals without resorting to
unitarity; this allows us to extract imaginary parts as well and gain
insight into the distributional structure of the integrands, as has
recently also been pointed out in \cite{Becher:2021urs} in the context
of an effective field theory analysis \cite{Becher:2016mmh}. The
physics of non-global observables has also been addressed starting
from a duality to small-$x$ evolution in \cite{Caron-Huot:2015bja},
and the resummation of non-global logarithms at the NLL level has also
been achieved by \cite{Banfi:2021owj,Banfi:2021xzn}. The purpose of
this work is to approach this resummation program, at the same level,
from the colour evolution point of view. We will derive the algorithm
we have previously been studying at the LL level
\cite{Martinez:2018ffw} that is implemented for full colour evolution
in the \texttt{CVolver} library
\cite{Platzer:2013fha,DeAngelis:2020rvq}, however we go beyond this in
three aspects: We will formulate the structure of the evolution at the
next order to incorporate the two loop ingredients
\cite{Platzer:2020lbr} and double soft currents
\cite{Catani:1999ss,Catani:2000pi}; we will point out how
hadronization can enter the calculation of cross sections, making
firmer links to our previous observation that part of the physics
modeled by colour reconnection and cluster fission might actually stem
from a perturbative interface to colour evolution
\cite{Gieseke:2017clv,Gieseke:2018gff}; and we will analyse how the
new evolution factor impacts the accuracy and attempt to relate our
analysis to factorization theorems to make contact with first
principle analyses and other approaches. Once our soft algorithm is
extended to the hard collinear case, a task which mainly consists of a
more complicated book keeping, we will be able to address more general
algorithms which then would account for the splitting functions we
have identified in \cite{Loschner:2021keu}. Definitions of singular
phase space regions become crucial to derive the evolution equations
and generalize the usual parton shower infrared cutoff. We show that
variations of the partonic shower cutoff necessarily force the soft
function (and thus a hadronization model) to compensate for changes in
the hard evolution. This might connect to previous and ongoing work on
the interpretation of the top mass parameter \cite{Hoang:2018zrp}, as
well as similar EFT approaches to hadronization corrections following
the concepts set out in \cite{Hoang:2007vb}. While we have now gained
a significant amount of understanding for the QCD case, electroweak
physics only now is recognized as in need for a treatment beyond the
customary high energy evolution addressing the quasi-collinear
limit. In recent work we have pointed out out that an amplitude
factorization similar to QCD can serve as a starting point for
evolution \cite{Platzer:2022nfu} and part of the present work's
motivation is to address electroweak evolution. This is particularly
tied to the question of what role the measurement and the projection
on the physical final state will play (also see \cite{Maas:2017wzi}
for a new conceptual approach to this question that highlights the
similarity with the need for hadronization in QCD).

\subsection{Aim of the present work and guide to the paper}

\begin{figure}
  \centering
  \includegraphics[scale=0.7]{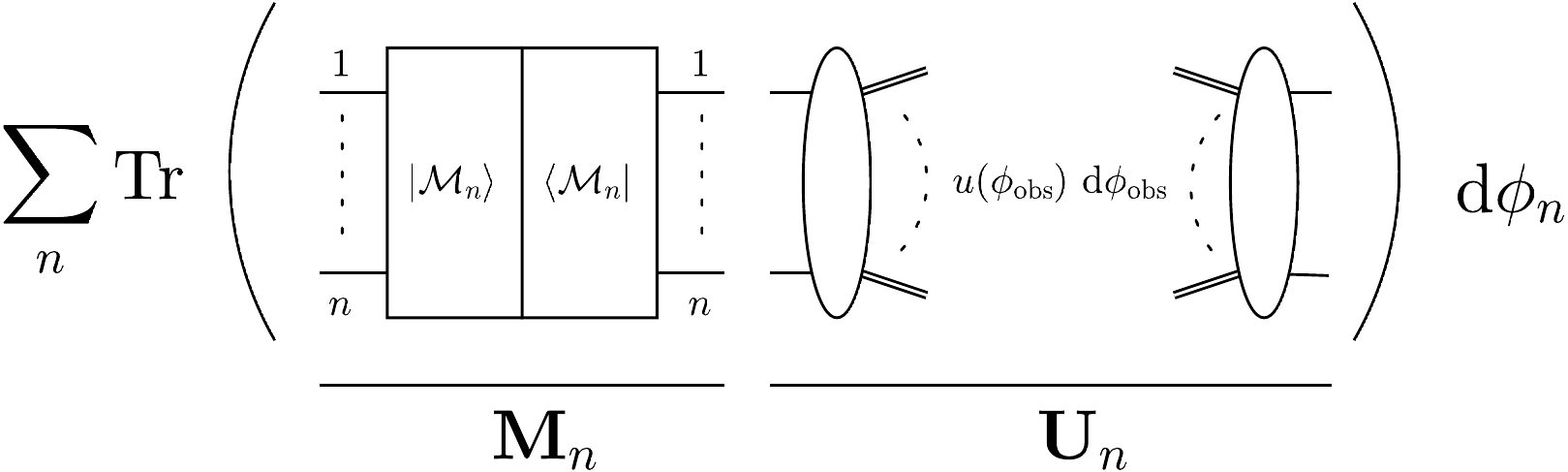}
  \caption{\label{fig:xsec}Graphical representation of the present
    approach.  A cross section is calculated by considering a partonic
    amplitude and its conjugate as an operator in colour space of
    intermediate partonic configurations. We sum over all
    configurations weighted by a projection on final state colour
    structures and a measurement, integrated over the finally observed
    objects. The details of the measurement are not of concern to the
    present work but we will investigate to what extent this
    factorization dictates evolution equations for both ${\mathbf
      M}_n$ and ${\mathbf U}_n$.}
\end{figure}

Our starting point will be a cross section definition which factors
hard scattering matrix elements from possibly non-perturbative factors
in the amplitude which describe the formation of hadrons and the
implementation of the observable of interest. We shall perform this
analysis on the level of ``density operators'' in colour space, as we
have previously studied for partonic cross sections in earlier work
(see {\it e.g.}
\cite{Martinez:2018ffw,Forshaw:2019ver,Platzer:2020lbr} for the
general formalism and to to set the notation). Graphically we
represent our approach in Fig.~\ref{fig:xsec}.
We assume that we will be able
to calculate the scattering matrix element and its conjugate,
conveniently assembled into the cross section density operator
${\mathbf M}_n = |{\cal M}_n\rangle\langle{\cal M}_n|$ in perturbation
theory and as a vector $|{\cal M}_n\rangle$ in the space of colour
structures for $n$ additionally emitted partons on top of a hard
scattering process. The cross section density operator, in a given
basis, would thus be represented as a quadratic matrix
\cite{Martinez:2018ffw}. The combined effect of projecting the
partonic final state on certain colour configurations, and on
performing a measurement described by an observable function on the
final state, will then be encoded in an operator ${\mathbf U}_n$,
which contains the truly quantum mechanical external states and
operators which overlap with these states in the sense that we can
create and annihilate (hard) partons from these states. The form of
the resulting cross section will be (schematically)
$$
  \sigma \sim \sum_n \int {\rm Tr}\left[{\mathbf M}_n {\mathbf U}_n\right]{\rm d}\phi_n
$$ where ${\rm d}\phi_n$ indicates an integration over the partonic
  momenta (effectively, the partonic phase space, though this is
  already a very specific assumption).  ${\mathbf U}_n$ might be
  structured such that it allows or prevents cancellation of infrared
  singularities. In the latter case we need to consider ${\mathbf
    U}_n$ as a genuinely non-perturbative object such as a
  fragmentation function or parton distribution function is. The aim of
  the present paper is not to work out the complete definition of
  ${\mathbf U}_n$ and the conditions under which we can actually write
  a cross section in the form above (we will give an outlook on this
  topic), but rather what the implications of the shear presence of a
  non-trivial ${\mathbf U}_n$ are: we analyse to what extent we can
  use the anticipated factorization to consistently construct
  evolution equations -- and this mainly means finite, iterative
  algorithms -- to build up the hard scattering operator and the
  measurement operator (which we can think of as anything in between a
  jet measurement, fragmentation function, or a completely exclusive
  hadronization model), subject to a given class of observables. Any
  finite algorithm, if it shall be based on events of fixed
  multiplicity, necessarily involves an infrared resolution. The
  presence of this resolution, {\it e.g.}  through an infrared cutoff
  scale $\mu_S$, shall be our starting point. We will devise
  re-definitions (or renormalization transformations) from ${\mathbf
    M}_n$ and ${\mathbf U}_n$,
$$
{\mathbf M}_n Z_g^n = {\cal Z}_n\left[{\mathbf A}(\mu_S),\mu_S\right]
\qquad {\mathbf U}_n = {\cal X}_n\left[{\mathbf S}(\mu_S),\mu_S\right]
$$ onto finite density operators ${\mathbf A}(\mu_S) = ({\mathbf
  A}_0(\mu_S),{\mathbf A}_1(\mu_S),...)$ and effective measurement
operators ${\mathbf S}(\mu_S) = ({\mathbf S}_0(\mu_S),{\mathbf
  S}_1(\mu_S),...)$ such that the cross section is invariant in the
sense that
\begin{multline}\nonumber
\sum_n \alpha_0^n \int {\rm Tr}\left[{\mathbf M}_n {\mathbf U}_n\right]{\rm d}\phi_n =
\sum_n \alpha_S^n \int {\rm Tr}\left[{\cal Z}_n\left[{\mathbf A}(\mu_S),\mu_S\right]
 {\cal X}_n\left[{\mathbf S}(\mu_S),\mu_S\right]  
  \right]{\rm d}\phi_n =\\
\sum_n \alpha_S^n \int {\rm Tr}\left[{\mathbf A}_n(\mu_S) {\mathbf S}_n(\mu_S)\right]{\rm d}\phi_n \ ,
\end{multline}
where $\alpha_0 = \alpha_S Z_g$ relates the bare and
  renormalized couplings, and $\mu_S$ is a resolution scale (in fact,
  as we demonstrate below, a collection of resolutions scales), of
  which the final cross sections needs to be independent in the same
  way as it is of the renormalization scale (which we have supressed
  for simplicity in these introductory notes). Jet cross sections
would be recovered if ${\mathbf S}_n$ are the unit operators in colour
space times a scalar measurement function, and would then be
represented by a scalar product matrix of overlaps of colour states,
see \cite{Martinez:2018ffw} for more details.  The independence of the
``bare'' objects ${\mathbf M}_n$ and ${\mathbf U}_n$ of the scale
$\mu_S$ will result in evolution equations for the ${\mathbf
  A}_n(\mu_S)$ and ${\mathbf S}_n(\mu_S)$, and also implies the fact
that the cross section is independent of the chosen resolution
scale. Had we chosen to work with vectors of density operators
${\mathbf A}$ rather their components ${\mathbf A}_n$ for each
partonic multiplicity, then the action of ${\mathcal Z}$ and
${\mathcal X}$ would in fact be matrices of colour charge operators
which are inverse to each other. While the former notation might look
more transparent on conceptual grounds, we here aim at a practical
analysis, which is better based on the density operators for
individual partonic multiplicities and the cross feed between them,
mediated by the emission of additional partons. We should thus warn
the reader that the implementations of the re-definitions ${\mathcal
  Z}$ and ${\mathcal X}$ in the main text will appear in a more
complicated fashion than by representing them as simple linear
operators acting on all the density operators. However we exactly
exploit the fact that ${\cal X}$ is inverse to ${\cal Z}$, and we
spell out and prove this relation in very detail in the main text: We
will actually determine ${\mathcal Z}$ from ${\cal X}$ to be its
inverse, possibly order-by-order in the strong coupling
$\alpha_S$. ${\cal X}$ itself needs to be determined to provide
infrared subtractions to the hard density operator, after ultraviolet
renormalization, and its factorization properties. This step very much
resembles what happens in a fixed-order calculation in which infrared
divergences are subtracted either by explicit subtraction terms or by
re-defining objects like a PDF or fragmentation function. In
particular this procedure needs to demonstrate how {\it iterated}, or
uncorrelated, unresolved emissions are removed from simultaneous
unresolved limits of two or more emissions, such as {\it e.g.}
discussed in the context of the work presented in
\cite{Banfi:2021owj,Banfi:2021xzn}. We explicitly demonstrate that our
formalism is able to achieve this and it thus gives rise to a finite
${\mathbf A}$, which is given by the correlated matrix elements at the
relevant orders without contaminating these by iterated contributions.

This rest of this work is organized as follows: In
Sec.~\ref{sec:infrared} we will start from the postulate of a
factorization theorem and show how the renormalization of a soft
factor gives rise to systematic infrared subtraction terms, thus
recovering the structure of a fixed-order calculation. From the very
fact that we want to eventually calculate a finite cross section we
deduce in Sec.~\ref{sec:resummation}, how the hard cross section
density matrix is factored into a finite contribution which accounts
for large logarithms, and renormalization factors which absorb the
universal infrared divergences exhibited by the subtraction terms. The
former object will satisfy the evolution equation recently proposed in
\cite{Martinez:2018ffw} and generalized to the hard collinear case in
\cite{Forshaw:2019ver}. This evolution equation, as well as the
evolution equation for the soft function will be discussed in detail
in Sec.~\ref{sec:evolution}, where we provide all necessary
ingredients up to the second order in $\alpha_S$. To show how we will
combine real and virtual corrections subject to a resolution criterion
driven by the class of observables we aim at, we give explicit
one-loop results in Sec.~\ref{sec:leading}, while a full application
of the two loop results will be saved for a follow-up publication. In
Sec.~\ref{sec:accuracy} we comment on the structure of the second
order evolution, and highlight how the colour structure and the
resolution criteria are constrained by a given observable.  In
Sec.~\ref{sec:hadronization} we finally comment on how the soft
function can also be viewed as a hadronization model including colour
reconnection, effectively generalizing the role fragmentation or
parton distribution functions would play in canceling out infrared
divergences.

\section{Infrared Subtractions and Measurement Operators}
\label{sec:infrared}

We will only focus on soft gluon evolution, and generalize the
starting point of the algorithm presented in \cite{Martinez:2018ffw}
to the following {\it Ansatz} by writing a cross section as
\begin{equation}
  \label{eq:factorization}
  \sigma[{\mathbf U}] = \sum_n \int \alpha_0^n\ {\rm Tr} \left[{\mathbf
      M}_n(Q;p_1,...,p_n) {\mathbf U}_n(Q;p_1,...,p_n)\right]  {\rm
    d}\phi(Q)\prod_{i=1}^n (4\pi\mu^2)^{\epsilon}[{\rm d}p_i]\tilde{\delta}(p_i) \ ,
\end{equation}
where we consider the emission of $n$ additional soft partons
on top of a set of hard momenta collectively referred to as $Q$. We
assume that the additional partons are ordered (in the simplest case
in energy), such that a full cross section might need to sum over
different permutations of orderings, but this does not make a
difference for our arguments below. Effectively, the
  ordering reflects a hierarchy upon which we can factorize the
  amplitudes. Specifically it does so in a way that if a certain
  number of last emissions with respect to this ordering cannot be
  singular through some phase space restriction, that any emission
  before will not lead to divergences of the same kind. Virtuality,
  transverse momenta or similar variables \cite{Nagy:2014mqa} might be
  used to achieve this, but also more complicated combinations of
  energy and angles can serve the same purpose. $\theta$ functions
implementing the ordering are thus always implicit when we refer to
additional emissions and the ordering corresponds to the labeling of
the momenta. $\alpha_0$ is the bare QCD coupling, and $\mu$ the 't
Hooft mass as we are working in dimensional regularization with
$d=4-2\epsilon$ dimensions.
\begin{equation}
  [{\rm d}p_i] = \frac{{\rm d}^dp_i}{\pi^{d/2}}\quad\text{ and }\quad \tilde{\delta}(p_i) = 2\pi \delta(p_i^2) \theta(p_i^0)
\end{equation}
denote the integration over final state momenta, but we will use the
same notation to denote integration over loop momenta, and
$\tilde{\delta}$ is the mass-shell restriction of the physical phase
space.\footnote{Notice that, in accordance with soft factorization, we
  assume that momentum conservation is strictly only applied to the
  hard momenta; however, more generally, we can always cast phase
  space factorization into the form given above and might contain
  final state momentum conservation in ${\mathbf U}$.}  In writing
this formula we have assumed a factorization of the cross section in a
sense that the measurement definition, and in particular infrared and
possibly non-perturbative phenomena are contained in ${\mathbf
  U}_n$. This operator can encode jet cross sections when we choose it
to be ${\mathbf 1}_n\times u(Q;p_1,...,p_n)$, the identity operator in
colour space times a scalar observable function. If we were to include
hard-collinear contributions we could also address cross sections for
identified hadrons in case of which the observable function would
constrain partonic momenta to be aligned along a certain direction. In
the latter case we would have assumed that there are further,
unobserved, hadrons for which the accompanying partonic activity would
provide the colour charges needed to neutralize colour of observed
hadronic final states. It is this mechanism which will break down if
we ask for a completely exclusive final state, or even for an
identified hadron with a specific quark content. In both of the latter
cases projections of colour into certain singlet systems is needed and
${\mathbf U}$, at any fixed partonic multiplicity, cannot be the
identity operator in colour space. This shall be our starting
point. In the conclusions, Sec~\ref{ref:outlook}, we will sketch how
this picture can arise from a first principle analysis, work which we
will defer in detail to an upcoming paper.

\subsection{Infrared subtractions for the hard density operator}

The hard density operator ${\mathbf M}_n$ is calculable in
perturbation theory as
\begin{equation}
  \label{eq:bareexpansion}
  {\mathbf M}_n = \sum_{l=0}^\infty\alpha_0^l\ {\mathbf M}_n^{(l)} \ ,
\end{equation}
where each term ${\mathbf M}_n^{(l)}$ refers to the product of an
amplitude and its conjugate (see
\cite{Martinez:2018ffw,Forshaw:2019ver,Platzer:2020lbr,Loschner:2021keu}
for more details and aspects of this formalism) for $n$ emissions, and
$l$ loops. The sum of the amplitude's and conjugate amplitude's number
of loops equals $l$ and it is understood that they are distributed
symmetrically. For example a one-loop term would read
\begin{equation}
  {\mathbf M}_n^{(1)} = |{\cal M}_n^{(0)}\rangle\langle {\cal M}_n^{(1)}| + |{\cal M}_n^{(1)}\rangle\langle {\cal M}_n^{(0)}|
\end{equation}
in terms of tree-level and one-loop amplitudes $|{\cal
  M}_n^{(0)}\rangle$ and $|{\cal M}_n^{(1)}\rangle$,
respectively. Using the renormalized coupling at a renormalization
scale $\mu_R$,
\begin{equation}
  \label{eq:runningcoupling}
    \alpha_0\left(4\pi\mu^2\right)^{\epsilon} = \alpha_S(\mu_R)\mu_R^{2\epsilon}Z_g\qquad Z_g= \sum_{l\ge 0}\alpha_S^l Z_g^{(l)}
\end{equation}
we will remove the ultraviolet divergences from ${\mathbf
  M}_n$\footnote{This procedure is of course only a net effect of a
  full renormalization program for the case of massless partons which
  we consider here. It is therefore also understood that none of the
  virtual corrections include explicit contributions from truncated
  legs, which we simply put to zero as scaleless integrals. More
  details on renormalization factors in the context of factorized
  amplitudes are explicitly discussed in \cite{Platzer:2022nfu} and
  can be adapted from there for more general cases.}. However, subject
to the structure of the observable, infrared divergences might
partially or fully cancel, or remain all together. We will therefore
start to individually remove these divergences with a subtraction
formalism which we express by re-defining the observable operator
as\footnote{We suppress the explicit dependence on the emission's
  momenta as far as no confusion arises; most of the time this will be
  the case since we implicitly assume that recoil is either irrelevant
  or can be accounted for by only transforming the hard system
  $Q$. For the case of soft gluon evolution this is mostly justified,
  and generalizations of our formalism to include the full complexity
  of hard-collinear physics should be straightforward and will be
  reported in future work.}
\begin{equation}
  \label{eq:renormobservable}
  {\mathbf U}_n = {\cal X}_n[{\mathbf S}(\mu_S),\mu_S] =  {\mathbf X}^\dagger_n {\mathbf S}_n {\mathbf X}_n -
  \sum_{s=1}^\infty \alpha_S^s \int{\mathbf F}_{n+s}^{(s)\dagger}
      {\mathbf S}_{n+s} {\mathbf F}_{n+s}^{(s)}
      \prod_{i=n+1}^{n+s}\mu_R^{2\epsilon}[{\rm
          d}p_i]\tilde{\delta}(p_i)\ .
\end{equation}
The dependence on $\mu_S$ appears in all of the quantities ${\mathbf
  S}_n$, ${\mathbf X}_n$ and ${\mathbf F}^{(s)\dagger}_{n+s}{\mathbf
  S}_{n+s} {\mathbf F}_{n+s}$ on the right hand side and has only been
supressed for readability. In the redefinition,
\begin{equation}
    {\mathbf X}_n = 1-\sum_{k\ge 1} \alpha_S^k {\mathbf
      X}_n^{(k)}\qquad\text{and}\qquad {\mathbf F}_n^{(s)\dagger} \circ {\mathbf F}_n^{(s)} =
    \sum_{k+\bar{k}\ge 0} \alpha_S^{k+\bar{k}} {\mathbf F}_n^{(s,k)\dagger}\circ {\mathbf F}_n^{(s,\bar{k})}
\end{equation}
will be used as infrared counter-terms for virtual corrections in the
$n$-parton amplitude, ${\mathbf X}_n$, and for $s$ unresolved partons
emitted from an $n$ parton state, ${\mathbf F}^{(s)}_n$. Here, the
$\circ$ denotes that the operator is understood to act on the left,
and on the right of a given operator in colour space: Subject to
fixing a particular basis in colour space (or possibly including other
quantum numbers as well \cite{Platzer:2022nfu}), the ${\mathbf X}_n$
operators mix colour basis structures (and would hence be represented
as quadratic matrices acting on the colour space for $n$ partons),
while emission operators such as ${\mathbf F}_n^{s}$ do change to a
larger parton ensemble, hence a larger basis of colour states and as
such would be represented as rectangular matrices.  In the case of
only soft gluons, or other means to entirely work at the level of
factorizing amplitudes \cite{Platzer:2022nfu}, there is in principle
no reason to link both operators, though we choose to do so here. We
will explicitly keep ${\mathbf X}_n$ and ${\mathbf
  F}_n^{(s)\dagger}\circ {\mathbf F}_n^{(s)}$ to not be related by
unitarity, since in presence of a definite projection on final states
virtual and real contributions might not relate one-to-one. In
Sec.~\ref{sec:observables} we discuss one example where the unitarity
assumption might lead to a problematic algorithm. In particular we
expect that this will be the case when considering electroweak
evolution when cuts through exchanges of vector bosons involve finite
width effects, see \cite{Platzer:2022nfu} for more details. In
general, we can only demand ${\mathbf S}$ to define the physical
observable for which we will be able to calculate a finite cross
section. However, if it is possible to demand that also ${\mathbf U}$
reflects a physical measurement, then we can use
Eq.~\ref{eq:renormobservable} to assess the accuracy of the
subtraction we perform.  Eq.~\ref{eq:renormobservable} reflects by
what amount the original definition of the measurement is altered with
respect to the re-defined one after subtraction, something which we
will be discussing in more detail in Sec.~\ref{sec:accuracy}. We will
in particular address this case in relation to hadronization
corrections for colour evolution and their connection to hadronization
models in Sec.~\ref{sec:hadronization}, though a generalization to
fragmentation functions and parton distribution functions will become
obvious by then, as well.

\begin{figure}
  \begin{center}
    \includegraphics[scale=0.5]{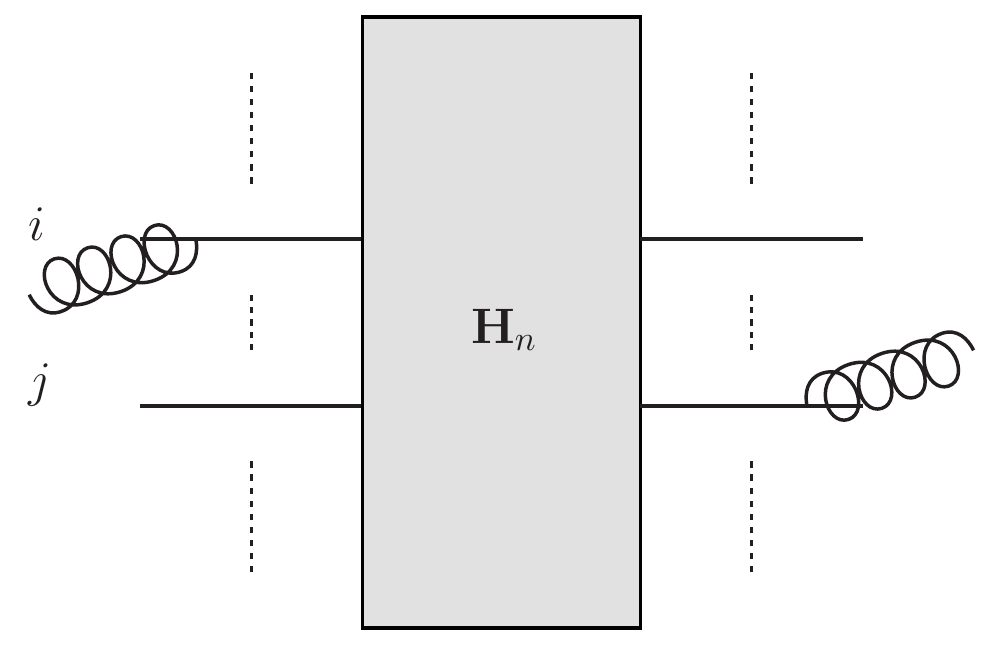}\includegraphics[scale=0.5]{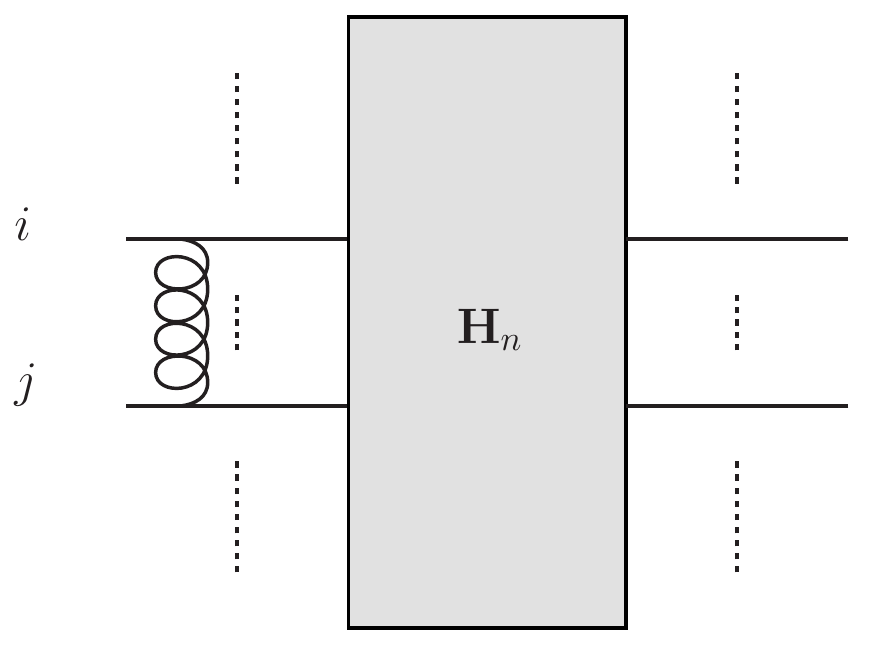}
    \includegraphics[scale=0.5]{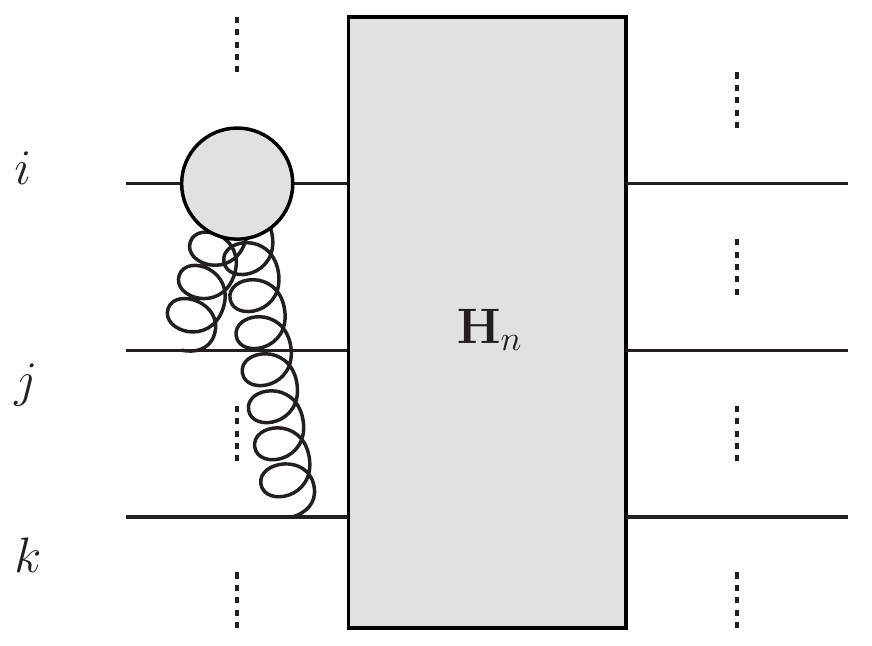}\includegraphics[scale=0.5]{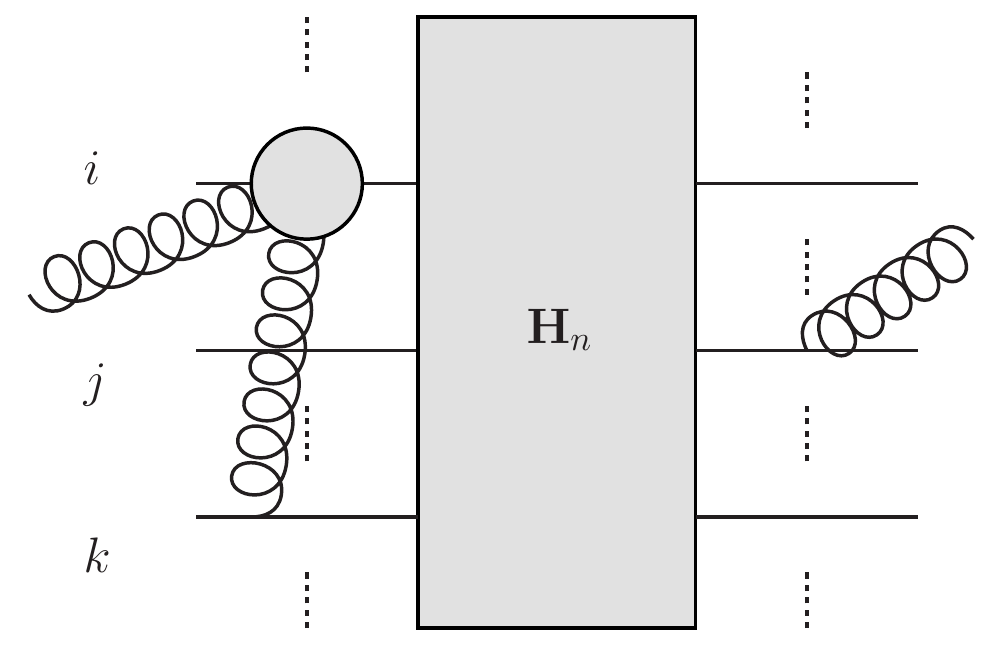}
    \includegraphics[scale=0.5]{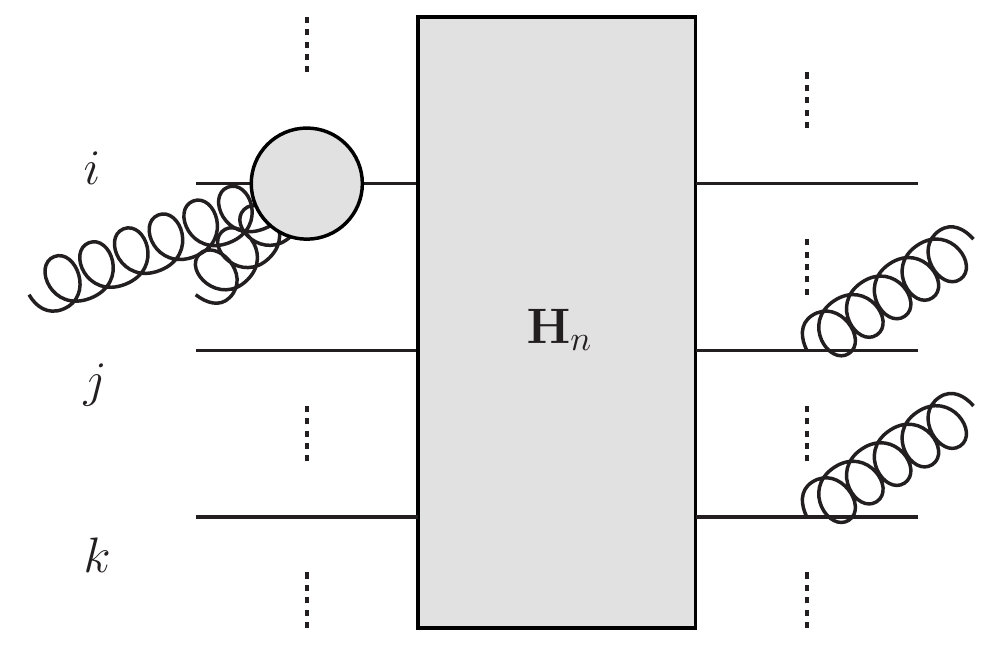}
  \end{center}
  \caption{\label{fig:doperators}Contributions to the different
    emission and virtual correction operators. From top left to bottom
    right we show graphs contributing to
    ${\mathbf D}_n^{(1,0)}\circ {\mathbf D}_n^{(1,0)\dagger}$,
    ${\mathbf V}_n^{(1)}\circ {\mathbf 1}$,
    ${\mathbf V}_n^{(2)}\circ {\mathbf 1}$,
    ${\mathbf D}_n^{(1,1)}\circ {\mathbf D}_n^{(1,0)\dagger}$ and
    ${\mathbf D}_n^{(2,0)}\circ {\mathbf D}_n^{(2,0)\dagger}$
    respectively. For further details see the main text. The reader is
    also referred to
    \cite{Catani:1999ss,Catani:2000pi,Platzer:2020lbr} for further
    reference in the context of soft gluons, and to
    \cite{Loschner:2021keu} for emission operators beyond this limit.}
\end{figure}

To see how the re-definition of the measurement
Eq.~\ref{eq:renormobservable} facilitates the task of providing
infrared subtractions we expand the cross section
Eq.~\ref{eq:factorization} in the renormalized coupling
Eq.~\ref{eq:runningcoupling}. This can of course be done at a fixed,
minimum, multiplicity, in case of which we reproduce the structure of
a fixed-order calculation. Anticipating that we do want to construct
an all-emission, all-orders cross section it is more instructive to
consider fixed order expansions at {\it each} multiplicity, {\it i.e.}
expanding up to including ${\cal O}(\alpha_S^{n+n_c})$ at each, fixed
$n\ge 0$. Once these contributions are rendered finite by the
subtraction, the all-order expression which we will derive in the next
section will be finite at all orders which iterate ${\cal
  O}(\alpha_S^{n_c})$ building blocks. We therefore consider to
re-write the cross section as
\begin{equation}
\label{eq:foexp}
  \left.\sigma[{\mathbf U}]\right|_{n_c} = \sum_{n \ge 0} \left( \int\alpha_S^n\ {\rm Tr} \left[\hat{{\mathbf
      M}}_n {\mathbf S}_n\right]  {\rm
    d}\phi(Q)\prod_{i=1}^n \mu_R^{2\epsilon}[{\rm d}p_i]\tilde{\delta}(p_i) +\ {\cal O}(\alpha_S^{n+n_{c}+1})\right) \ ,
\end{equation}
where ${\hat{\mathbf M}}_n = \sum_{l\ge 0}\alpha_S^l {\hat{\mathbf
    M}}_n^{(l)}$. The individual $\hat{\mathbf M}_n^{(l)}$ can then be
found by comparing fixed orders at fixed multiplicities, where we only
consider $l\le n_c$ and $s+l\le n_c$ for the expansion of the
operators ${\mathbf X}_n^{(l)}$ and ${\mathbf F}_n^{(s,l)}$,
respectively. In particular we find for $n_c=2$
\begin{eqnarray}
  \hat{{\mathbf M}}_n^{(0)} &= &{\mathbf M}_{n}^{(0)} - {\cal S}^{(2)}_{\text{tree}}\left[{\mathbf M}_{n}^{(0)}\right] \\
  \hat{{\mathbf M}}_n^{(1)} &= &{\mathbf M}_{n,R}^{(1)}- {\cal S}^{(2)}_{\text{tree}}\left[{\mathbf M}_{n,R}^{(1)}\right]-
      {\cal S}^{(2)}_{1\text{-loop}}\left[{\mathbf M}_{n,R}^{(0)}\right]\\
\hat{{\mathbf M}}_n^{(2)} &= &{\mathbf M}_{n,R}^{(2)}- {\cal S}^{(2)}_{\text{tree}}\left[{\mathbf M}_{n,R}^{(2)}\right]
- {\cal S}^{(2)}_{1\text{-loop}}\left[{\mathbf M}_{n,R}^{(1)}\right] -
{\cal S}^{(2)}_{2\text{-loop}}\left[{\mathbf M}_{n,R}^{(0)}\right]
\end{eqnarray}
where ${\mathbf M}_{n,R}^{(0)} = {\mathbf M}_{n}^{(0)}$,
\begin{equation}
  {\mathbf M}_{n,R}^{(1)} =   {\mathbf M}_{n}^{(1)} + n Z_g^{(1)}   {\mathbf M}_{n}^{(0)}
\end{equation}
and
\begin{equation}
  {\mathbf M}_{n,R}^{(2)} =   {\mathbf M}_{n}^{(2)} + (n+1) Z_g^{(1)}   {\mathbf M}_{n}^{(1)}
  + \left(\frac{n(n-1)}{2} \left(Z_g^{(1)}\right)^2 + n Z_g^{(2)}\right){\mathbf M}_{n}^{(0)}
\end{equation}
define the renormalized density operators ${\mathbf M}_{n,R}^{(l)}$
with UV finite virtual corrections up to two loops, and it is
understood that ${\mathbf M}_{n}^{(l)}$ is zero whenever $n<0$ or
$l<0$. The operators ${\cal S}^{n_c}_L$ (with
$L=\text{tree},\text{1-loop},...)$ act on specific contributions
${\mathbf H}_n^{(l)}\equiv {\mathbf M}_{n,R}^{(l)}$ to the hard
density operator with $n$ legs and $l$ loops to provide infrared
subtractions required up to an expansion to ${\cal O}(\alpha_S^{n_c})$
in the sense of Eq.~\ref{eq:foexp}. The result of ${\cal
  S}_L^{(n_c)}\left[{\mathbf H}_n^{(l)}\right]$ involves other density
operator contributions ${\mathbf H}_{m}^{(k)}$ with $k\le l$ loops
and/or $m\le n$ legs, but the colour structure of ${\cal
  S}_l^{(n_c)}\left[{\mathbf H}_n^{(l)}\right]$ is the one pertaining
to a density operator of $n$ partons. Had we chosen to collect density
operators per partonic multiplicity in an infinite vector, then ${\cal
  S}$ would be a matrix which mixes different multiplicities though we
refrain from such a presentation since the change between different
partonic multiplicities is an important concept in the problems we
analyze in the present work. It is also important to note that the
$\hat{\mathbf M}_n$ are now finite through the subtractions at each
multiplicity, but {\it only} up to the additional orders
$\alpha_S^{n_c}$ we have considered in the present
analysis. Information about higher orders will only be available once
we have facilitated the re-definition of ${\mathbf M}_n$ to ${\mathbf
  A}_n$ thus completing the factorization and renormalization program
for the cross section. To this end we have distinguished $\hat{\mathbf
  M}_n$ from ${\mathbf A}_n$, though they will coincide if the
$\hat{\mathbf M}_n$ have effectively been build up by the evolution
equation to be put in place for ${\mathbf A}_n$ later. We in
particular find that
\begin{equation}
  {\cal S}^{(2)}_{\text{tree}}\left[{\mathbf H}^{(l)}_n\right] =
  {\mathbf F}_n^{(1,0)} {\mathbf H}^{(l)}_{n-1} {\mathbf F}_n^{(1,0)\dagger} +
    {\mathbf F}_n^{(2,0)} {\mathbf H}^{(l)}_{n-2} {\mathbf F}_n^{(2,0)\dagger}
\end{equation}
provides subtractions for unresolved limits stemming from tree-level
emission diagrams, for which we can choose ${\mathbf F}_{n}^{(1,0)}$
to be given by the single soft emission current $\hat{\mathbf
  D}_{n}^{(1,0)}$ (see Ap.~\ref{sec:oversubtract} for more details).
Notice that this is an over-subtraction in the double unresolved
limit, where ${\mathbf H}_{n-1}^{(l)}$ admits factorization of another
emission, as explicitly analyzed in Ap.~\ref{sec:oversubtract}. We
therefore need to take
\begin{eqnarray}
    {\mathbf F}_n^{(1,0)} \circ {\mathbf F}_n^{(1,0)\dagger} &= &\hat{\mathbf D}_{n}^{(1,0)} \circ \hat{\mathbf
      D}_{n}^{(1,0)\dagger}
    \Theta_{n,1}\\ {\mathbf F}_n^{(2,0)} \circ {\mathbf
      F}_n^{(2,0)\dagger} &= &  \hat{\mathbf D}_n^{(2,0)} \circ \hat{\mathbf
      D}_n^{(2,0)\dagger}\Theta_{n,2} - \hat{\mathbf D}_n^{(1,0)}\hat{\mathbf
      D}_{n-1}^{(1,0)} \circ \hat{\mathbf D}_{n-1}^{(1,0)\dagger}\hat{\mathbf
      D}_n^{(1,0)\dagger}\Theta_{n,1}
\end{eqnarray}
in terms of the true unresolved behaviour $\hat{\mathbf D}_n^{(2,0)}$
as {\it e.g.}  given by the soft gluon currents \cite{Catani:1999ss}.
Diagramatically, these contributions have been depicted in
Fig.~\ref{fig:doperators}, among some of their real/virtual or purely
virtual counter parts, which will be dicsussed later in the text.
We also require a resolution criterion $\Theta$ which
projects onto singular regions: $\Theta_{n,k}$ is supposed to be one
for {\it any} singular configuration of $k$ out of $n$ partons (and
zero outside), and this in particular implies that
${\cal S}^{(2)}_{\text{tree}}[{\mathbf H}_n^{(l)}]$ vanishes whenever
$n$ partons are fully resolved. Let us illustrate the
  requirements of these resolution criteria for example when cutting
  on energies and angles of the emitted partons. We would then choose
\begin{equation}
\Theta_{n,1} = 1-\hat{\Theta}_{n,1}\theta(E_n-\mu_S)
\end{equation}
or, for two emissions,
\begin{equation}
\Theta_{n,2} =  1-\hat{\Theta}_{n,2}\theta(E_{n-1}-\mu_S)\theta(E_{n}-\mu_S) \ ,
\end{equation}
where in both cases $\hat{\Theta}_{n,1}$ and $\hat{\Theta}_{n,2}$
indicate resolved angular separations: they will vanish on those phase
space regions which allow for collinear divergences, {\it e.g.} one could choose
\begin{equation}
  \hat{\Theta}_{n,1} = \prod_{i=1}^{n-1} \theta\left(\frac{p_i\cdot p_n}{E_i E_n} - \lambda\right)
\end{equation}
however all of these choices need to be carefully subjected to
analysing the classes of observables we intent to reliably
predict. This is further detailed in Sec.~\ref{sec:accuracy}.  Also
notice that we strictly have no need to use step functions. Had we
chosen resolution in virtualities, then one choice of resolution
function could be
\begin{equation}
\Theta_{n,1} = 1-  \prod_{i=1}^{n-1} \theta\left(s_{in} - \mu_S^2\right)
\end{equation}
for one emission (where $s_{ij}=(p_i+p_j)^2$), and possible
generalizations for more than one emission. Again, the resolution
function would be unity whenever any propagator would become
singular. Cutting in virtuality for more than one emission is somewhat
more complicated, though the requirement that $\Theta_{n,2}$ becomes
one whenever any of the singular limits is reached, indicates that
also here a product over cuts on all virtualities of the emissions
would be sufficient. In what we have presented so far, the
$\Theta_{n,s}$ functions have been applied as global factors to the
emission operators, or as global insertions into the integrands of
virtual matrix elements (see below). There is, however, no need to
limit this to this case and one can, as well, understand this
prescription to act differently for different contributions (even
individual graphs like those depicted in
Fig.~\ref{fig:doperators}). As an example, in the case of virtuality
cuts, we could group the contributions in terms of divergencies
involving a certain set of invariants, say those resembling the
behaviour of the fifth graph in \ref{fig:doperators} would cut on
\begin{equation}
  \Theta_{n,2}^{\text{(graph 5 topologies)}} = 1 -
  \prod_{s = s_{i,n},s_{i,n-1},s_{n,n-1},s_{jn},s_{k,n-1}}\theta(s-\mu_S^2)
\end{equation}
had we chosen in this case to attach the gluon $n-1$ to the line $k$
on the right hand side of the diagram. Generally,
$\hat{\mathbf D}_{n}^{(s,l)}$ originates from factorizing soft gluon
emissions from the {\it renormalized} hard density matrix operators
and is thus free of UV divergences (see App.~\ref{sec:oversubtract}
for a precise definition of what factorization properties we assume
here). For loop integrals this in particular implies that their
genuine UV divergences have been removed and the corresponding
integrations are cut off accordingly (more aspects of this will be
highlighted in future work). We further discuss the removal of
over-subtraction in App.~\ref{sec:oversubtract}, and stress the fact
the subtraction terms are constructed such that the subtracted
quantities are entirely projected on those phase space regions which
are free of singularities. We might split up the action of the
emission operators into different terms and use different $\Theta$
functions {\it e.g.}  depending on the dipole in between a gluon is
exchanged, though we will not make this explicit here for the sake of
readability. Furthermore
\begin{equation}
  {\cal S}^{(2)}_{1\text{-loop}}\left[{\mathbf H}^{(l)}_n\right] =
  {\mathbf X}_n^{(1)} {\mathbf H}^{(l)}_n+{\mathbf H}^{(l)}_n {\mathbf X}_n^{(1)\dagger}
  +{\mathbf F}_n^{(1,1)} {\mathbf H}^{(l)}_{n-1} {\mathbf F}_n^{(1,0)\dagger} +
  {\mathbf F}_n^{(1,0)} {\mathbf H}^{(l)}_{n-1} {\mathbf F}_n^{(1,1)\dagger}
\end{equation}
provides one-loop subtractions, and
\begin{equation}
  {\cal S}^{(2)}_{2\text{-loop}}\left[{\mathbf H}_{n}^{(l)}\right]
  = {\mathbf X}_n^{(2)} {\mathbf H}^{(l)}_n+{\mathbf H}^{(l)}_n {\mathbf X}_n^{(2)\dagger}
  -{\mathbf X}_n^{(1)} {\mathbf H}^{(l)}_n {\mathbf X}_n^{(1)\dagger}
\end{equation}
provides two-loop subtractions.  Similarly to what happened in the
emission case, we need to adjust ${\mathbf X}_n^{(1)}$ to be given by
the singular (or unresolved) part of the one-loop exchange
$\hat{\mathbf V}_n^{(1)}[\Xi_{n,1}]$, where $\Xi_{n,k}$ refers to a
restriction on $k$ unresolved loop momentum modes in presence of $n$
other resolved momenta,
\begin{equation}
  {\mathbf X}_n^{(1)} = \hat{\mathbf V}_n^{(1)}[\Xi_{n,1}] \left(1-\xi \Theta_{n,1}\right) +
\hat{\mathbf V}_n^{(1)}\xi  \Theta_{n,1} \ ,
\end{equation}
with the shorthand $\hat{\mathbf V}_n^{(l)}\equiv \hat{\mathbf
  V}_n^{(l)}[1]$. $\xi =0,1$ labels two alternatives of the
subtraction, and both yield a finite algorithm. Their difference
amounts to whether the matrix element from which virtual exchanges
have been factors is rendered finite for emissions, as well ($\xi=1$)
or not ($\xi=0$). We can therefore think of $\xi=0$ amounting to an
inclusive algorithm, for which we would allow the observable function
to be infrared safe and allow for unresolved radiation, and $\xi=1$
would refer to a fully exclusive algorithm in which we will not rely
on any cancellation in between different partonic multiplicities. The
latter effectively means that we can achieve a merging of different
fixed-order calculations, though we will not explore the consequences
further in the present work.  More details are given in
App.~\ref{sec:oversubtract}.  Following \cite{Platzer:2020lbr}, we use
a sum over cuts $\alpha$ and write the virtual exchanges on external
lines as
\begin{equation}
  \hat{\mathbf V}_n^{(l)}[\Xi_{n,l}] =\sum_\alpha \int {\mathbf {\cal
    I}}_{n,\alpha}^{(l)}(p_1,...,p_n;k_1,...,k_l)\ \Xi^{(\alpha)}_{n,l} \prod_{i=1}^l
  \mu_R^{2\epsilon} [{\rm d}k_i] \ ,
\end{equation}
where $\Xi_{n,l}^{(\alpha)}$ is the analogue of $\Theta_{n,k}$ for the
virtual corrections (though we, again, might split this up into more
terms, or combine cuts with each other). Since any more complicated
situation follows by generalization we again keep things simple and
refer to one function $\Xi_{n,l}$ in the following. ${\cal
  I}_{n,\alpha}^{(l)}(p_1,...,p_n;k_1,...,k_l)$ denotes the integrand,
which might include $\delta$-functions which pin down the loop momenta
to particular kinematic regions. They thus are distributions in
general, including the $i0$ prescriptions.  For the two-loop case we
use 
\begin{equation}
{\mathbf X}_n^{(2)} = \hat{\mathbf
  V}_n^{(2)}[\Xi_{n,2}](1-\xi\Theta_{n,2}) + \hat{\mathbf V}_n^{(2)} \xi \Theta_{n,2}
-{\mathbf X}_n^{(1)}\hat{\mathbf V}_n^{(1)}
\end{equation}
in terms of the singular parts of the two-loop exchanges
$\hat{\mathbf V}_n^{(2)}$ (note that the
${\mathbf X}_n^{(1)}\circ {\mathbf X}_n^{(1)\dagger}$ subtractions
automatically adjust for the two loop case). These virtual corrections
again stem from factorization of the renormalized matrix elements, and
it is thus understood that suitable UV counter terms have been added
to their definition, or that the genuine UV divergent integrals are
cut off accordingly (artificial UV divergences might arise from the
soft gluon approximation and are understood to be cut off by the
resolution functions). Some examples are again given in
Fig.~\ref{fig:doperators}. {\it E.g.}  the one-loop self energy
insertion on a one-loop soft gluon exchange would require a UV counter
term which resembles $Z_g^{(1)}$ upon integration (such counter terms
can readily be constructed by using the methods developed in {\it
  e.g.}  \cite{Becker:2010ng,Capatti:2022tit}). No other genuine UV
divergences arise in the two-loop case.  Constructing counter terms
for the one-loop/one-emission contributions which are free of
over-subtractions we find
\begin{eqnarray}
  {\mathbf F}_n^{(1,1)}\circ {\mathbf F}^{(1,0)\dagger}_n &=&
 \hat{\mathbf D}_n^{(1,1)} \circ \hat{\mathbf D}^{(1,0)\dagger}_n \Theta_{n,1}\\\nonumber
&+&  \hat{\mathbf D}_n^{(1,1)}\left[\Xi_{n,1}\right]
  \circ \hat{\mathbf D}^{(1,0)\dagger}_n(1- \Theta_{n,1})\\\nonumber
  &-&{\mathbf X}_n^{(1)} \hat{\mathbf D}_n^{(1,0)}  \circ \hat{\mathbf D}^{(1,0)\dagger}_n
  - \hat{\mathbf
    D}_n^{(1,0)}\hat{\mathbf V}_{n-1}^{(1)}
  \circ \hat{\mathbf D}^{(1,0)\dagger}_n\Theta_{n,1}
\end{eqnarray}
in terms of the full unresolved behaviour $\hat{\mathbf D}_n^{(1,1)}$
(and similarly for
${\mathbf F}_n^{(1,0)}\circ {\mathbf F}^{(1,1)\dagger}_n$), which does
not exactly coincide with the one-loop soft gluon current of
\cite{Catani:2000pi} which in turn has some iterated contributions
already removed. However it can easily be constructed from this
result. On the very same note $\hat{\mathbf V}_n^{(2)}$ and
$\hat{\mathbf D}_n^{(2,0)}$ are understood to contain the full sum of
contributing diagrams, see App.~\ref{sec:oversubtract} for details,
and Fig.~\ref{fig:doperators} for sample diagrams.

\section{Resummation}
\label{sec:resummation}

In the regions of unresolved emissions and exchanges we expect the
hard density operator to factorize as
\begin{equation}
  \label{eq:renormdensity}
  {\mathbf M}_n Z_g^n = {\cal Z}_n[{\mathbf A}(\mu_S),\mu_S] = {\mathbf Z}_n {\mathbf A}_n {\mathbf Z}_n^\dagger+
  \sum_{s=1}^n \alpha_S^s {\mathbf E}_n^{(s)} {\mathbf A}_{n-s} {\mathbf E}_n^{(s)\dagger}
\end{equation}
where
  \begin{equation}
  {\mathbf Z}_n = 1+\sum_{k\ge 1} \alpha_S^k {\mathbf Z}_n^{(k)}\qquad
{\mathbf E}_n^{(s)}\circ {\mathbf E}_n^{(s)\dagger}  = \sum_{k+\bar{k}\ge 0} \alpha_S^{k+\bar{k}} {\mathbf E}_n^{(s,k)}\circ {\mathbf E}_n^{(s,\bar{k})\dagger} \ .
  \end{equation}
It is important to stress that this re-definition is to be thought of
in the sense that we renormalize the divergent density operator
${\mathbf M}_n$ and trade it off for a finite, though possibly
logarithmically enhanced, object ${\mathbf A}_n$ from which we can
hope to build up an algorithm for the calculation of physical,
infrared finite, cross sections. This operation needs to constitute
the inverse of the transformation providing infrared subtractions such
that the cross section stays invariant under the combined
transformation. We thus will be able to infer the operators ${\mathbf
  Z}_n$ and ${\mathbf E}_n$ from their counterparts ${\mathbf X}_n$
and ${\mathbf F}_n$. Also note that the left hand side of
Eq.~\ref{eq:renormdensity} is ultraviolet finite, such that we solely
accumulate infrared divergences in the renormalization factors on the
right hand side. We can then perform fixed-order expansions of
${\mathbf M}$ using Eq.~\ref{eq:bareexpansion}, and
\begin{equation}
  {\mathbf A}_n = \sum_{l=0}^\infty \alpha_s^l {\mathbf A}_n^{(l)} \ .
\end{equation}
The cross section reads
\begin{equation}
  \sigma[{\mathbf U}] = \sum_n \int \alpha_S^n\ {\rm Tr} \left[\left({\mathbf
      A}_n  + {\mathbf \Delta}_{n}\right){\mathbf S}_n   \right] {\rm
    d}\phi(Q)\prod_{i=1}^n \mu_R^{2\epsilon}[{\rm d}p_i]\tilde{\delta}(p_i) \ .
\end{equation}
where $\Delta_{n}$ encodes the effects beyond the exact cancellation
of the combined redefinitions Eq.~\ref{eq:renormobservable} and
Eq.~\ref{eq:renormdensity}, and Eq.~\ref{eq:runningcoupling}.  We can
enforce ${\mathbf \Delta}_n = 0$, if 
\begin{multline}
  \label{eq:inverses}
  {\mathbf Z}_n = {\mathbf X}_n^{-1}\quad \text{ and }\\
  {\mathbf X}_n {\mathbf E}_{n}^{(s)} \circ  {\mathbf E}_{n}^{(s)\dagger}{\mathbf X}_n^\dagger
  - {\mathbf F}_n^{(s)} {\mathbf Z}_{n-s}\circ {\mathbf Z}_{n-s}^\dagger {\mathbf F}_n^{(s)\dagger} -
  \sum_{t=1}^{s-1} {\mathbf F}_n^{(t)} {\mathbf E}_{n-t}^{(s-t)}\circ {\mathbf E}_{n-t}^{(s-t)\dagger} {\mathbf F}_n^{(t)\dagger} = 0 
\end{multline}
holds to all orders in $\alpha_S$. This is the previously mentioned
invariance of the cross section under the combined transformation of
the hard scattering and measurement operators: transforming ${\mathbf
  M}_n$ into ${\mathbf A}_n$ is precisely {\it inverse} to
transforming ${\mathbf U}_n$ into ${\mathbf S}_n$. To this extend note
that, since the trace is cyclic, the cross section involves a sum over
all mutliplicities, and phase space factorizes, the action of
re-defining ${\mathbf U}_n$ in terms of ${\mathbf S}_n$ can be recast
into an action on the ${\mathbf M}_n$ as
\begin{equation}\nonumber
  {\mathbf M}_n Z_g^n {\mathbf U}_n = \left[{\mathbf X}_n{\mathbf M}_n Z_g^n {\mathbf X}^\dagger_n - \sum_{s=1}^n\alpha_S^s
   {\mathbf F}_{n}^{(s)}{\mathbf M}_{n-s} Z_g^{n-s} {\mathbf F}_{n}^{(s)\dagger}\right] {\mathbf S}_n 
\end{equation}
where equality in the above holds upon performing the trace, the
summation over all mutliplicities and the integration over phase
space. Had we chosen to arrange all ${\mathbf M}_n$ and all ${\mathbf
  U}_n$ each into a vector with elements indexed by partonic
multiplicity, the above would then be a linear operation, though
several of the key features of the evolution will then again be
hidden. Inserting the re-definition of ${\mathbf M}_n Z_g^n$ directly
shows that the operations are inverse to each other so long
Eq.~\ref{eq:inverses} holds.  In the below we will determine ${\mathbf
  Z}$ and ${\mathbf E}$ in this way order by order, however the
formula above might serve as a more general starting point and proves
to be especially relevant in the electroweak case, in which ${\mathbf
  F}$ might involve decay processes only (corresponding to observed
final states), while ${\mathbf E}$ might involve both, emission of
unstable particles which subsequently count as hard lines, as well as
their decays, and we need to match these identities at the level of
different (final) states rather than at fixed order.

In QCD, at fixed order, we simply adjust the renormalization constants
as
\begin{equation}
  {\mathbf Z}_n^{(1)} = {\mathbf X}_n^{(1)}\qquad {\mathbf
    E}_n^{(1,0)}\circ {\mathbf
    E}_n^{(1,0)\dagger}  = {\mathbf F}_n^{(1,0)}\circ  {\mathbf F}_n^{(1,0)\dagger} \ ,
\end{equation}
then ${\mathbf \Delta}_n$ will contribute at ${\cal
  O}(\alpha_S^{n+2})$ at fixed ${\mathbf S}_n$, and with ${\mathbf Z}_n^{(2)} = {\mathbf X}_n^{(2)} + \left({\mathbf X}_n^{(1)}\right)^2$,
\begin{equation}
  {\mathbf E}_n^{(1,1)}\circ {\mathbf E}_n^{(1,0)} = {\mathbf F}_n^{(1,1)}\circ {\mathbf F}_n^{(1,0)}
  + {\mathbf X}_n^{(1)} {\mathbf F}_n^{(1,0)}\circ {\mathbf F}_n^{(1,0)}
  + {\mathbf F}_n^{(1,0)}{\mathbf X}_{n-1}^{(1)}\circ {\mathbf F}_n^{(1,0)}
\end{equation}
and
\begin{equation}
  {\mathbf E}_{n}^{(2,0)}\circ {\mathbf E}_n^{(2,0)\dagger} =
  {\mathbf F}_{n}^{(2,0)}\circ {\mathbf F}_n^{(2,0)\dagger}
  + {\mathbf F}_{n}^{(1,0)}{\mathbf F}_{n-1}^{(1,0)}\circ {\mathbf F}_{n-1}^{(1,0)\dagger}{\mathbf F}_n^{(1,0)\dagger}
\end{equation}
${\mathbf \Delta}_n$ will contribute at ${\cal O}(\alpha_S^{n+3})$ at
fixed ${\mathbf S}_n$. Together with the removal of the
over-subtractions we then obtain
\begin{eqnarray}
  {\mathbf Z}_n^{(2)} &=&
  \hat{\mathbf V}_n^{(2)}[\Xi_{n,2}](1-\xi\Theta_{n,2})-
  \hat{\mathbf V}_n^{(1)}[\Xi_{n,1}]\hat{\mathbf V}_n^{(1)}[1-\Xi_{n,1}](1-\xi\Theta_{n,1})^2\\\nonumber
  &+& \hat{\mathbf V}_n^{(2)}\xi \Theta_{n,2} - \xi \Theta_{n,1}(1-\xi\Theta_{n,1})\hat{\mathbf V}_n^{(1)}
  \hat{\mathbf V}_n^{(1)}[1-\Xi_{n,1}]
  \ ,
\end{eqnarray}
\begin{eqnarray}
  {\mathbf E}_n^{(1,1)}\circ {\mathbf E}^{(1,0)\dagger}_n &=&
 \hat{\mathbf D}_n^{(1,1)} \circ \hat{\mathbf D}^{(1,0)\dagger}_n \Theta_{n,1}\\\nonumber
&+&  \hat{\mathbf D}_n^{(1,1)}\left[\Xi_{n,1}\right]
  \circ \hat{\mathbf D}^{(1,0)\dagger}_n(1- \Theta_{n,1})\\\nonumber
&-&  \hat{\mathbf
    V}_n^{(1)}\left[\Xi_{n,1}\right] \hat{\mathbf D}_n^{(1,0)}\circ \hat{\mathbf D}^{(1,0)\dagger}_n(1-\Theta_{n,1})(1-\xi\Theta_{n,1}) \\\nonumber
&-& \hat{\mathbf
    D}_n^{(1,0)}\hat{\mathbf V}_{n-1}^{(1)}\left[1-\Xi_{n-1,1}\right]
  \circ \hat{\mathbf D}^{(1,0)\dagger}_n\Theta_{n,1}(1-\xi\Theta_{n-1,1})\\\nonumber
  &-&\hat{\mathbf V}_n^{(1)} \hat{\mathbf D}_{n}^{(1,0)}\circ \hat{\mathbf D}_{n}^{(1,0)\dagger}(1-\Theta_{n,1}) \xi \Theta_{n,1}
\end{eqnarray}
and
\begin{equation}
  {\mathbf E}_{n}^{(2,0)}\circ {\mathbf E}_n^{(2,0)\dagger} =
  \hat{\mathbf D}_n^{(2,0)} \circ \hat{\mathbf D}_n^{(2,0)\dagger}\Theta_{n,2} -
  \hat{\mathbf D}_n^{(1,0)}\hat{\mathbf D}_{n-1}^{(1,0)} \circ \hat{\mathbf D}_{n-1}^{(1,0)\dagger} \hat{\mathbf D}_n^{(1,0)\dagger}
  \Theta_{n,1}(1-\Theta_{n-1,1}) \ .
\end{equation}
Note that we have left expressions for $\Theta_{n,1}(1-\Theta_{n,1})$
or $\Theta_{n,1}^2$ to fully keep track of the distributional nature,
noting that {\it e.g.}  $\theta(x)(1-\theta(x))= 0$ everywhere except
$x=0$, while still $({\rm d}/{\rm d}x)\theta(x)(1-\theta(x)) = 0$, and
also of course $1-\theta(x) + \theta(x) = 1$. Other choices of
resolution functions, however, might thus also be possible to use.

Let us stress the importance of the over-subtraction removal, {\it
  e.g.} in the case of the real emission. Iterating the redefinition
of ${\mathbf M}$ for two emissions from a hard configuration with no
emissions, we would have
\begin{eqnarray}
{\mathbf M}^{(0)}_2 &=&
 {\mathbf A}^{(0)}_2 + 
{\mathbf D}^{(1,0)}_{1}{\mathbf A}^{(0)}_1 {\mathbf D}^{(1,0)\dagger}_{1}\Theta_{2,1}\\\nonumber& +&
\left({\mathbf D}^{(2,0)}_{0}{\mathbf H} {\mathbf D}^{(2,0)\dagger}_{0}\Theta_{2,2}
- \textcolor{blue}{ {\mathbf D}^{(1,0)}_{1} {\mathbf D}^{(1,0)}_{0}{\mathbf H} {\mathbf D}^{(1,0)\dagger}_{0}
{\mathbf D}^{(1,0)\dagger}_{1}\Theta_{2,1}(1-\Theta_{1,1}) } \right)
\end{eqnarray}
as well as
\begin{equation}
{\mathbf A}^{(0)}_1 = {\mathbf M}^{(0)}_1-
{\mathbf D}^{(1,0)}_{0}{\mathbf H} {\mathbf D}^{(1,0)\dagger}_{0}\Theta_{1,1} \ ,
\end{equation}
where we have put ${\mathbf A}^{(0)}_0 = {\mathbf H}$. Clearly,
${\mathbf A}^{(0)}_1$ is manifestly finite in the singly-unresolved
limit where ${\mathbf M}^{(0)}_1 \to {\mathbf D}^{(1,0)}_{0}{\mathbf
  H} {\mathbf D}^{(1,0)\dagger}_{0}$. We also have
\begin{eqnarray}
{\mathbf A}^{(0)}_2 &=&
 {\mathbf M}^{(0)}_2 -
{\mathbf D}^{(1,0)}_{1} \left({\mathbf M}^{(0)}_1-
{\mathbf D}^{(1,0)}_{0}{\mathbf H} {\mathbf D}^{(1,0)\dagger}_{0}\Theta_{1,1}
\right)    {\mathbf D}^{(1,0)\dagger}_{1}\Theta_{2,1}\\\nonumber& -&
\left({\mathbf D}^{(2,0)}_{0}{\mathbf H} {\mathbf D}^{(2,0)\dagger}_{0}\Theta_{2,2}
- \textcolor{blue}{ {\mathbf D}^{(1,0)}_{1} {\mathbf D}^{(1,0)}_{0}{\mathbf H} {\mathbf D}^{(1,0)\dagger}_{0}
{\mathbf D}^{(1,0)\dagger}_{1}\Theta_{2,1}(1-\Theta_{1,1}) } \right)
\end{eqnarray}
In the doubly unresolved limit both $ {\mathbf M}^{(0)}_2\to {\mathbf
  D}^{(2,0)}_{0}{\mathbf H} {\mathbf D}^{(2,0)\dagger}_{0}$ and
${\mathbf M}^{(0)}_1 \to {\mathbf D}^{(1,0)}_{0}{\mathbf H} {\mathbf
  D}^{(1,0)\dagger}_{0}$. Hence the above gives rise to
\begin{eqnarray}
{\mathbf A}^{(0)}_2& \to &  {\mathbf D}^{(2,0)}_{0}{\mathbf H} {\mathbf D}^{(2,0)\dagger}_{0}(1-\Theta_{2,2})\\\nonumber
&-& {\mathbf D}^{(1,0)}_{1} {\mathbf D}^{(1,0)}_{0}{\mathbf H} {\mathbf D}^{(1,0)\dagger}_{0}
{\mathbf D}^{(1,0)\dagger}_{1}\Theta_{2,1}(1-\Theta_{1,1})\\\nonumber
&\textcolor{blue}{+}& \textcolor{blue}{ {\mathbf D}^{(1,0)}_{1} {\mathbf D}^{(1,0)}_{0}{\mathbf H} {\mathbf D}^{(1,0)\dagger}_{0}
{\mathbf D}^{(1,0)\dagger}_{1}\Theta_{2,1}(1-\Theta_{1,1}) } \ .
\end{eqnarray}
Thus, both the presence and the choice of resolution function on the
blue contribution are needed such that ${\mathbf A}_2^{(0)}$ is finite
(even in a singly unresolved limit when $\Theta_{2,2}\to 1$ and
$\Theta_{2,1}\to 1$) and solely -- without double counting -- given by
the regularized double-emission factors and finite in all doubly
unresolved limits. No finite algorithm could be achieved had we left
out the blue contribution. It is in particular also the
presence of the blue contribution which guarantees that in the ordered
phase space regions iterated contributions are removed, while we need
to consider the full matrix elements elsewhere (to be discussed in
more details in the next section). Similar remarks apply to the other
contributions, and in Ap.~\ref{sec:oversubtract} we give a detailed
analysis of this construction.

\section{Evolution Equations}
\label{sec:evolution}

We will now study how the dependence on the different infrared
resolution scales as well as the renormalization scale $\mu_R$ gives
rise to evolution equations for the hard and soft factors.  Notice
that the running of $\alpha_S$ is given by the $d$-dimensional
$\beta$-function as
\begin{equation}
  \partial_R \alpha_S \equiv \mu_R\frac{\partial
    \alpha_S}{\partial\mu_R} = \alpha_S \sum_{k\ge 0} \beta_{R,k-1}
  \alpha_S^k \equiv \beta(\alpha_S)\alpha_S \qquad \beta_{R,-1} = -2\epsilon \ ,
\end{equation}
and it is understood that $\beta_{S,k}=0$ if $S\ne R$ indicates that
we differentiate to a different scale $\mu_S$ in place of the
renormalization scale $\mu_R$. From Eq.~\ref{eq:runningcoupling} we
then have at second order
\begin{equation}
  \beta_{R,0} = 2\epsilon Z_g^{(1)} \qquad \beta_{R,1} = 4\epsilon \left(Z_g^{(2)} - (Z_g^{(1)})^2\right)
\end{equation}

\subsection{Hard density operator}

If we differentiate Eq.~\ref{eq:renormdensity} to any of the
resolution scales $\mu_S$ contained in $\Theta_{n,s}$ and $\Xi_{n,s}$,
or the renormalization scale, we then obtain ($\partial_S \equiv
\partial/\partial \log \mu_S$ as above)
\begin{equation}
  \partial_S {\mathbf A}_n = {\mathbf \Gamma}_{n,S}{\mathbf A}_n + {\mathbf A}_n {\mathbf \Gamma}_{n,S}^\dagger -
  \sum_{s\ge 1} \alpha_S^s {\mathbf R}_{S,n}^{(s)} {\mathbf A}_{n-s} {\mathbf R}_{S,n}^{(s)\dagger}
\end{equation}
where
\begin{equation}
  {\mathbf \Gamma}_{n,S} = -Z_g^{\frac{n}{2}} \partial_S Z_g^{-\frac{n}{2}} - {\mathbf Z}_n^{-1}\partial_S {\mathbf Z}_n
\end{equation}
and
\begin{eqnarray}
  {\mathbf R}_{S,n}^{(s)} \circ {\mathbf R}_{S,n}^{(s)\dagger} &=& {\mathbf Z}_n^{-1}\left[
  \left(Z_g^{n}\partial_S Z_g^{-n} +s \ \beta(\alpha_S)\right)
  {\mathbf E}_{n}^{(s)} \circ {\mathbf E}_{n}^{(s)\dagger}\right.\\\nonumber &+&
  \partial_S\left({\mathbf E}_{n}^{(s)} \circ {\mathbf E}_{n}^{(s)\dagger}\right)  +
          {\mathbf E}_{n}^{(s)} {\mathbf \Gamma}_{n-s,S}\circ {\mathbf E}_{n}^{(s)\dagger}
          + {\mathbf E}_{n}^{(s)}\circ {\mathbf \Gamma}_{n-s,S}^{\dagger} {\mathbf E}_{n}^{(s)\dagger}
  \\\nonumber
  &-&\left.
  \sum_{s'=1}^{s-1}{\mathbf E}_{n}^{(s-s')} {\mathbf R}_{S,n-s+s'}^{(s')} \circ {\mathbf R}_{S,n-s+s'}^{(s')\dagger} {\mathbf E}_{n}^{(s-s')\dagger}
  \right] \left({\mathbf Z}_n^{\dagger}\right)^{-1}\ .
\end{eqnarray}
Explicit expressions up to second order are obtained in
App.~\ref{sec:anomhard}. Note that the appearance of $\beta_{S,-1}$
will mostly be canceled by derivatives acting on the $\mu_R$ factors,
thus no explicit account of their effect will be needed. For the cases
where an explicit account of it is needed we employ (in a
distributional sense for functions with support $x\in (0,1]$)
\begin{equation}
  \frac{\epsilon}{x^{1-\epsilon}} = \delta(x) - \sum_{n=0}^\infty \frac{\epsilon^{n+1}}{n!} \left[\frac{\ln(1/x)}{x}\right]_+ \ ,
\end{equation}
where the $+$ prescription is to be taken with a subtraction at
$x=0$. Notice however, that most of such cases will actually be
dropping out due to the $d$-dimensional $\beta$ function.

\subsection{Soft function}

The transformed soft function is subject to an evolution
\begin{equation}
  \partial_S {\mathbf S}_n = -\tilde{\mathbf \Gamma}_{S,n}^\dagger {\mathbf S}_n - {\mathbf S}_n \tilde{\mathbf \Gamma}_{S,n}
  + \sum_{s\ge 1}\alpha_S^s
  \int \tilde{\mathbf R}_{S,n+s}^{(s)\dagger} {\mathbf S}_{n+s} \tilde{\mathbf R}_{S,n+s}^{(s)}
  \prod_{i=n+1}^{n+s}[{\rm d}p_i]\tilde{\delta}(p_i)
\end{equation}
where we now find the anomalous dimensions and evolution kernels to be
given by
\begin{equation}
  \tilde{\mathbf \Gamma}_{S,n} = \left(\partial_S {\mathbf X}_n\right) {\mathbf X}_n^{-1}
\end{equation}
and
\begin{eqnarray}
  \mu_R^{-2\epsilon s}\tilde{\mathbf R}_{S,n+s}^{(s)\dagger} \circ \tilde{\mathbf R}_{S,n+s}^{(s)} & = & \left({\mathbf X}_n^{\dagger}\right)^{-1}\left[
    s\ (\beta_S(\alpha_S)-\beta_{S,-1})
    {\mathbf F}_{n+s}^{(s)\dagger}\circ{\mathbf F}_{n+s}^{(s)}\right.\\\nonumber
    & + & \partial_S\left({\mathbf F}_{n+s}^{(s)\dagger}\circ{\mathbf F}_{n+s}^{(s)}\right)
             - {\mathbf F}_{n+s}^{(s)\dagger} \tilde{\mathbf \Gamma}_{n+s,S}^{\dagger}\circ {\mathbf F}_{n+s}^{(s)}
             - {\mathbf F}_{n+s}^{(s)\dagger}\circ \tilde{\mathbf \Gamma}_{n+s,S} {\mathbf E}_{n}^{(s)} \\\nonumber
             &+& \left.\sum_{s'=1}^{s-1}
             {\mathbf F}_{n+s-s'}^{(s-s')\dagger}
             \tilde{\mathbf R}_{S,n+s}^{(s')\dagger}\circ \tilde{\mathbf R}_{S,n+s}^{(s')} {\mathbf F}_{n+s-s'}^{(s-s')}\mu_R^{-2\epsilon s'}
             \right] {\mathbf X}_n^{-1} \ .
\end{eqnarray}
Explicit expressions up to second order are again given in
App.~\ref{sec:anomsoft}.  A final remark is in order with respect to
the choices of $\xi$. For $\xi=1$, we will obtain the anomalous
dimensions we would have used for $\xi=0$, accompanied by additional
factors which guarantee that the emissions in the state which the
anomalous dimension acts on, are all resolved. The second contribution
then is a purely finite virtual contribution acting by pinning down
the emissions to the resolution cutoff by means of a derivative of the
additional resolution function. The latter contribution thus gives a
specific contribution to configurations at the infrared cutoff, {\it
  i.e.} those which do not radiate any further. We will investigate
these specific aspects in future work, but in the following only
consider the case $\xi=0$, which also leads to a finite evolution.  As
a particular example, consider the one loop anomalous dimension (the
real emission at this order is not affected by the different choices
of $\xi$). Here we find, considering a soft evolution variable with
$\beta_S=0$ only,
\begin{equation}
  {\mathbf \Gamma}_S^{(1)} = -\hat{\mathbf V}_n^{(1)}\left[\partial_S \Xi_{n,1}\right](1-\xi \Theta_{n,1})
  - \hat{\mathbf V}_n^{(1)}\left[1-\Xi_{n,1}\right] \xi\ \partial_S \Theta_{n,1} \ .
\end{equation}

\section{Examples for Evolution at Leading Order}
\label{sec:leading}

In this section we consider the resummation of non-global logarithms
in an ordering with respect to energy. Specifically we require virtual
counter terms which can be inferred from the cuts obtained with the
methods in \cite{Platzer:2020lbr}. The one-loop virtual correction is
given by
\begin{multline}
  \mathbf{V}^{(1)}_n =
  \left(4\pi \mu^2\right)^{\epsilon}\ \frac{1}{2} \sum_{i < j}\left(-{\mathbf T}_i\cdot {\mathbf T}_j\right) \ \times\\
   \int
  \frac{[{\rm d} p_{n+1}]}{2\pi i} \frac{4 p_i\cdot p_j}
       {(2p_i\cdot p_{n+1}+i 0 (T\cdot p_{i})^2)(2p_j\cdot p_{n+1}-i 0 (T\cdot p_{j})^2)}
       \frac{1}{p_{n+1}^2+i 0 (T\cdot p_{n+1})^2}
\end{multline}
and admits the cuts given in \cite{Platzer:2020lbr}, out of which the
double cut vanishes in dimensional regularization and a real-valued
and imaginary single cut remains, if $i$ and $j$ are both incoming or
both outgoing, and the absorptive cut is canceled in the case of an
incoming/outgoing pair. These give rise to the subtractions (notice
that cutting introduces a minus sign in the use of the Feynman tree
theorem)
\begin{equation}
  \mathbf{X}^{(1)}_{n,\text{rad}} =
 \sum_{i < j}{\mathbf T}_i\cdot {\mathbf T}_j
  \int_0^\infty \frac{{\rm d}E}{E}\left(\frac{\mu_R}{E}\right)^{2\epsilon}
  \int_{-1}^{1} {\rm d}x
  \frac{(1-x^2)^{-\epsilon}}{1-x^2}
  \int {\rm d}\Omega^{(d-3)}\  \hat{\Xi}^{(ij)}_{n,1,\text{rad}}
\end{equation}
\begin{multline}
  \mathbf{X}^{(1)}_{n,\text{abs}} =
  \sum_{i < j}{\mathbf T}_i\cdot {\mathbf T}_j   \int_0^\infty {\rm d}E   \left(\frac{\mu_R}{E}\right)^{2\epsilon}\\
 \int_{-1}^{1} {\rm d}x
  (1-x^2)^{-\epsilon} \frac{1}{2 E x - i 0\ \sqrt{2p_i\cdot
      p_j}}\frac{1}{x^2-1+i 0\ 2 x^2}\ \hat{\Xi}^{(ij)}_{n,1,\text{abs}}
  \int {\rm d}\Omega^{(d-3)} \ .
\end{multline}
In this case we have assumed that $\xi=0$, and indicated that we have
evaluated these integrals in the dipole rest frame, which introduces
additional complications when combining with the real emission. In the
dipole frame, and for the radiative cut, we have $x = (p_i-p_j)\cdot
p_{n+1} / \sqrt{2 p_i\cdot p_j}$ and $E = (p_i+p_j)\cdot p_{n+1} /
\sqrt{2 p_i\cdot p_j}$ with $p_{n+1}^2=0$, and writing the integral in
the above form requires $\hat{\Xi}^{(ij)}_{n,1,\text{rad}}$ to be
symmetric around $x=0$. For the absorptive part, $E x = p_j\cdot
p_{n+1} / \sqrt{2 p_i\cdot p_j}$ and $p_i\cdot p_{n+1} = 0$. A boost
back to lab frame needs to separately map the dipole's legs into each
other, $\Lambda_{ij} \hat{n}_{i,j} = n_{i,j}$, where $p_{i,j}=E_{i,j}
n_{i,j}$, such that the cancellation with the real emission can be
ensured. In the case of a cutoff on energies we tag the exchange as
resolved, whenever its energy is within the real emission phase space
(here indicated by the limit on $Q$), or it has a minimum energy or a
minimum opening angle to one dipole leg,
\begin{equation}
\hat{\Xi}^{(ij)}_{n,1,\text{rad}} = 1- \theta(Q-E)\theta(E-\mu_S)\hat{\Theta}^{(ij)}_\lambda(1-x, 1+x) \ .
\end{equation}
Then we find for the anomalous dimensions from the radiative cut by
differentiating with respect to $\mu_S\partial/\partial \mu_S$ and
$\partial/\partial \lambda$
\begin{multline}
{\mathbf \Gamma}_{S,n,\text{rad}}^{(1)}= \tilde{{\mathbf \Gamma}}_{S,n,\text{rad}}^{(1)} = \\
- \sum_{i < j}{\mathbf T}_i\cdot {\mathbf T}_j\ \left(\frac{\mu_R}{\mu_S}\right)^{2\epsilon}\theta(Q-\mu_S)
\int_{-1}^{1} {\rm d}x
  \frac{(1-x^2)^{-\epsilon}}{1-x^2} \hat{\Theta}^{(ij)}_\lambda(1-x, 1+x) \int{\rm d}\Omega^{(d-3)}
\end{multline}
\begin{multline}
  \label{eq:cutoffad}
{\mathbf \Gamma}_{\lambda,n,\text{rad}}^{(1)} = \tilde{\mathbf \Gamma}_{\lambda,n,\text{rad}}^{(1)} = \sum_{i < j}{\mathbf
  T}_i\cdot {\mathbf
  T}_j\ \frac{1}{\epsilon}\left[\left(\frac{\mu_R}{Q}\right)^{2\epsilon} - \left(\frac{\mu_R}{\mu_S}\right)^{2\epsilon}\right]
\\\times  \int_{-1}^{1}
{\rm d}x \frac{(1-x^2)^{-\epsilon}}{1-x^2}
\left(-\frac{\partial}{\partial\lambda}\hat{\Theta}^{(ij)}_\lambda(1-x,
1+x)\right)  \int{\rm d}\Omega^{(d-3)}\ .
\end{multline}
Notice that the anomalous dimension for the angular cutoff would
exhibit a pole in $\epsilon$ if it was not for the upper bound on the
energy. We strongly suggest that this upper bound should not only be
inferred by considerations about the real cancellation, but by
amplitude-level considerations about the ordering variable as
advocated in
\cite{Angeles-Martinez:2015rna,AngelesMartinez:2016iow,Martinez:2018ffw}.
The role of the boost and frame dependence of the collinear cutoff
will be discussed in a separate work, notice only that
\begin{multline}
  {\cal F}(n_i\cdot n_j, n_i\cdot n, n_j\cdot n) {\rm d}\Omega^{(d-2)} =\\
  {\cal F}\left(2, \frac{1-x}{(\Lambda_{ij} \hat{n})^0}, \frac{1+x}{(\Lambda_{ij} \hat{n})^0}\right)
  \frac{1}{((\Lambda_{ij} \hat{n})^0)^{2-d}} {\rm d}x (1-x^2)^{-\epsilon} {\rm d}\Omega^{(d-3)}\\
  \equiv \hat{\cal F}^{(ij)}(1-x,1+x) {\rm d}x (1-x^2)^{-\epsilon} {\rm d}\Omega^{(d-3)}
\end{multline}
allows to translate between the lab frame and the frame in which we
consider regularizing the virtuals. The UV running at first order is
given by
\begin{equation}
{\mathbf \Gamma}_{R,n,\text{rad}}^{(1)} =\frac{n}{2}\beta_{R,0}\qquad\text{and}\qquad  \tilde{\mathbf \Gamma}_{R,n,\text{rad}}^{(1)} = 0 \ ,
\end{equation}
confirming the role of the hard and soft density operators. The
absorptive cut is regularized through the $i0$ terms which do not
cancel, and is integrable for any resolution which is symmetric
around, and not vanishing at $x=0$ (in fact, an angular cutoff is not
needed), provided that the energy is less than another `Glauber' scale
$\mu_G$:
\begin{equation}
  {\mathbf \Gamma}_{G,n,\text{rad}}^{(1)}= -\tilde{{\mathbf \Gamma}}_{G,n,\text{rad}}^{(1)} =
  \sum_{i<j} {\mathbf T}_i\cdot {\mathbf T}_j \frac{i\pi}{2}  \left(\frac{\mu_R}{\mu_G}\right)^{2\epsilon}\delta_{i,j\ \text{FF/II}}\int {\rm d}\Omega^{(d-3)} \ .
\end{equation}
The real emission subtraction would be given by the square of the
single soft gluon current,
\begin{equation}
{\mathbf E}_n^{(1,0)} \circ {\mathbf E}_n^{(1,0)\dagger} = {\mathbf F}_n^{(1,0)} \circ {\mathbf F}_n^{(1,0)\dagger} = 
-\sum_{i<j}{\mathbf T}_i \circ {\mathbf T}_j^\dagger \frac{p_i\cdot p_j}{p_i\cdot p_{n+1}\ p_{n+1}\cdot p_j}\Xi^{(ij)}_{n,1,\text{rad}}
\end{equation}
and accordingly be supplemented by a resolution criterion which we
take to be
\begin{equation}
  \Xi_{n,1,\text{rad}} = 1- \theta(p_{n+1}^0-\mu_S)\Theta_\lambda(n_i\cdot p_{n+1}/p_{n+1}^0, n_j\cdot p_{n+1}/p_{n+1}^0)
\end{equation}
and the derivatives analogously give the contributions to the
evolution from the real emissions. The evolution in the energy scale
then provides us with the soft gluon algorithm studied in
\cite{Martinez:2018ffw}; the evolution direction with the cutoff has
not yet been considered completely but will likely cancel for
observables in which collinear divergences cancel. We will leave this
to future work.\footnote{It should also be noted that
  \cite{Nagy:2022xae,Nagy:2022bph} have been reporting on similar
  multi-scale evolution algorithms while this work has been
  finalized. They only consider leading order evolution though and do
  not include the possible effects of hadronization as discussed later
  in this work.}

\section{Accuracy Considerations}
\label{sec:accuracy}

\subsection{Evolution at the Second Order and Comparison to Shower Approaches}

In this section we compare the structure of the evolution at the second
order to approaches of parton shower algorithms which aim to describe
evolution at the same accuracy. The results here can also be used to
assemble a full evolution using our results \cite{Platzer:2020lbr} and
the double emission current \cite{Catani:1999ss}. Putting
$\beta_{S,i}=0$ for the evolution of the hard density operator we
obtain
\begin{equation}
  {\mathbf \Gamma}_{S,n}^{(2)} = -\hat{\mathbf V}_{n}^{(2)}\left[\partial_S \Xi_{n,2}\right]
  + \hat{\mathbf V}_{n}^{(1)}\left[\partial_S \Xi_{n,1}\right]\hat{\mathbf V}_{n}^{(1)}\left[1- \Xi_{n,1}\right] \ ,
\end{equation}
\begin{eqnarray}
  \label{eq:1l1ehard}
  {\mathbf R}_n^{(1,1)}\circ {\mathbf R}_n^{(1,0)\dagger} &=&
  \left(\hat{\mathbf D}^{(1,1)}_n\left[1-\Xi_{n,1}\right] -
  \hat{\mathbf D}^{(1,0)}_n\hat{\mathbf V}_{n-1}^{(1)}\left[1- \Xi_{n-1,1}\right]\right)
  \circ \hat{\mathbf D}^{(1,0)\dagger}_n\partial_S \Theta_{n,1} \\\nonumber
  &+& \left(
  \hat{\mathbf D}^{(1,1)}_n\left[\partial_S\Xi_{n,1}\right] -
  \hat{\mathbf V}_{n}^{(1)}\left[\partial_S\Xi_{n,1}\right]\hat{\mathbf D}^{(1,0)}_n
  \right)\circ \hat{\mathbf D}^{(1,0)\dagger}_n (1-\Theta_{n,1}) \ ,
\end{eqnarray}
and
\begin{eqnarray}    \label{eq:2ehard}
  {\mathbf R}_{n}^{(2,0)}\circ {\mathbf R}_{n}^{(2,0)\dagger}
  &=& \hat{\mathbf D}_{n}^{(2,0)}\circ \hat{\mathbf D}_{n}^{(2,0)\dagger}\partial_S \Theta_{n,2}\\\nonumber
  &-& \hat{\mathbf D}_{n}^{(1,0)}\hat{\mathbf D}_{n-1}^{(1,0)}\circ \hat{\mathbf D}_{n-1}^{(1,0)\dagger}\hat{\mathbf D}_{n}^{(1,0)\dagger}
  (1-\Theta_{n-1,1})\partial_S \Theta_{n,1} \ .
\end{eqnarray}
Notice that both the double real, as well as the double virtual terms
have a form which resemble a subtraction calculation in the sense that
the ``earlier'' exchange or emission $n-1$ constrained to be resolved
by $1-\Xi_{n,1}$ or $1-\Theta_{n-1,1}$ and the later one is pinned
down to the evolution variable. Also Eq.~\ref{eq:1l1ehard} shows a
similar structure, and in fact the first contribution removes the
one-emission current after a virtual insertion from the
one-loop/one-emission contribution and so is in one-to-one
correspondence with directly using the result of \cite{Catani:2000pi},
while the second term corresponds to a different subtraction for the
correction of an emission which is resolved. It would thus be
interesting to compare the structure of such subtractions with
NLO-type subtraction scheme to ${\cal O}(\alpha_S^2)$ corrections of a
splitting kernel as advocated in \cite{Li:2016yez,Hoche:2017iem}. Let
us in particular emphasize how the resolution conspires with the
subtraction in the cases above, {\it e.g.} for the real emission. In
this case we can take
$\Theta_{n,1}=1-\hat{\Theta}_{n,1}\theta(E_n-\mu_S)$ with some angular
resolution $\hat{\Theta}_{n,1}$ which regulates collinear
divergences. In order to guarantee that the double emission phase
space is projected onto the singular region, we need to take
$\Theta_{n,2}=1-\hat{\Theta}_{n,2}\theta(E_{n-1}-\mu_S)\theta(E_{n}-\mu_S)$. Inserting
this into Eq.~\ref{eq:2ehard} we find that
  \begin{eqnarray}
    {\mathbf R}_{n}^{(2,0)}\circ {\mathbf R}_{n}^{(2,0)\dagger} &=&
    \left( \hat{\mathbf D}^{(0,2)}_{n}\circ \hat{\mathbf
      D}^{(0,2)\dagger}_{n}\hat{\Theta}_{n,2} - \hat{\mathbf
      D}^{(0,1)}_{n}\hat{\mathbf D}^{(0,1)}_{n-1}\circ \hat{\mathbf
      D}^{(0,1)\dagger}_{n-1}\hat{\mathbf
      D}^{(0,1)\dagger}_{n}\hat{\Theta}_{n-1,1}\hat{\Theta}_{n,1}\right)\\\nonumber&&\times \theta(E_{n-1}-\mu_S)\delta(E_n-\mu_s)\\\nonumber
    &+& \hat{\mathbf D}^{(0,2)}_{n}\circ \hat{\mathbf
      D}^{(0,2)\dagger}_{n}\hat{\Theta}_{n,2} \theta(E_n-\mu_S)\delta(E_{n-1}-\mu_S) \ ,
\end{eqnarray}
{\it i.e.} each of the two terms is completely regulated by the
angular and energy cutoffs, and subtractions of the iterated piece are
applied in the ordered region only.

\subsection{Accuracy for Various Observables}
\label{sec:observables}

The introduction of infrared subtractions provides us with a measure
of accuracy, and possibly the structure of hadronization corrections
which we briefly discuss in the next section. At the lowest order
Eq.~\ref{eq:renormobservable} implies that

\begin{multline}
  {\mathbf U}_n = {\mathbf S}_n
  - \alpha_S {\mathbf X}^{(1)\dagger}_n {\mathbf S}_n
  - \alpha_S {\mathbf S}_n {\mathbf X}^{(1)}_n\\
  -
 \alpha_S \int{\mathbf F}_{n+1}^{(1,0)\dagger}
      {\mathbf S}_{n+1} {\mathbf F}_{n+1}^{(1,0)}
      \mu_R^{2\epsilon}[{\rm d}p_{n+1}]\tilde{\delta}(p_{n+1})
      +{\cal O}(\alpha_s^2)\ .
\end{multline}
Up to this order we have a one-to-one correspondence between
subtraction terms and virtual and real corrections, see
Sec.~\ref{sec:infrared}.  Notice that even if we assume unitarity in
connecting $\hat{\mathbf V}_n^{(1)}$ and $\hat{\mathbf
  D}_n^{(1,0)\dagger}\circ \hat{\mathbf D}_n^{(1,0)}$ in the sense that we
choose $\Xi=\Theta$ and
\begin{equation}
\int\hat{\mathbf D}_{n+1}^{(1,0)\dagger}\hat{\mathbf D}_{n+1}^{(1,0)}\Theta_{n,1}
\mu_R^{2\epsilon}[{\rm d}p_{n+1}]\tilde{\delta}(p_{n+1}) =
-\frac{1}{2}\hat{\mathbf V}_n^{(1)}\left[\Theta_{n,1}\right]
\end{equation}
we will still not be able to obtain a simple relation for corrections
from the subtractions and observable, not even if only collinear
singularities are present since the emission operators in this case
collapse to colour diagonal contributions only for the very last
emission which allows to use the cyclicity of the trace. We stress
that even counter terms for Glauber exchanges will {\it not} drop out
of the evolution if ${\mathbf S}_n$ has non-trivial colour structure
(notice that there is no reason to assume that ${\mathbf S}_n$ will
trigger colour conservation, only ${\mathbf A}_n$ needs to). This
analysis highlights that we do need to consider a different structure
of power corrections once colour projections are involved in the
definition of a cross section. In particular we anticipate that this
not only happens for hadronization but also to the case of analyzing
electroweak exchanges at a similar level, using the ingredients
presented in \cite{Platzer:2022nfu}. The same remark applies to
initial state evolution which would appear in our framework alongside
the final state measurements and we assume that this will shed further
light on the physics of super-leading logarithms addressed in
\cite{Forshaw:2021fxs,Becher:2021zkk}.

Jet cross sections are the special case in which we would either start
from ${\mathbf U}_n = {\mathbf 1}_n u(p_1,...,p_n)$, or demand
${\mathbf S}_n={\mathbf 1}_n u(p_1,...,p_n)$. By the redefinition of
${\mathbf U}_n$ the cross section would then be re-arranged
accordingly, where the difference to the original cross section in
terms of ${\mathbf U}_n$ would have been encoded in the soft factor
${\mathbf S}_n$. If we demand that ${\mathbf S}_{n} = {\mathbf 1}_n
u(p_1,...,p_n)$ such as to reproduce the previous algorithms in
\cite{Martinez:2018ffw}, then we can invert
\begin{multline}
  {\mathbf U}_n = {\mathbf 1}_n u(p_1,...,p_{n})\\ -
  \alpha_s \int \mu_R^{2\epsilon}[{\rm
      d}p_{n+1}]\tilde{\delta}(p_{n+1}) \hat{\mathbf
    D}_{n+1}^{(1,0)\dagger}\hat{\mathbf D}_{n+1}^{(1,0)}\Theta_{n,1}
  \left[u(p_1,...,p_n,p_{n+1}) - u(p_1,...,p_{n})\right] + {\cal
    O}(\alpha_s^2)
\end{multline}
reproducing the genuine (power suppressed) difference in the
observables, {\it i.e.} essentially we would then be implementing a
slicing method. Notice that, in the case for the jet cross section,
the procedure above is exactly the same as adding the ${\cal
  O}(\alpha_S)$ term to a calculations defined in terms of ${\mathbf
  U}_n$ only to provide infrared subtractions compatible with the
behaviour of the observable.  In this case, the resolution criterion
$\Theta$ must be accordingly synchronized with the observable in
question.  In the case of a prototypical non-global observable we can
write
\begin{equation}
u(p_1,...,p_n,p_{n+1}) = (\theta(\rho-p^0_{n+1}) +
\theta(p^0_{n+1}-\rho) \theta(n_{n+1}\in \text{in}))u(p_1,...,p_n) \ ,
\end{equation}
and the quantity to analyze is thus
\begin{multline}
\Theta_{n,1}\left(u(p_1,...,p_n,p_{n+1})-u(p_1,...,p_n)\right) =
\left(1-\theta(p_{n+1}^0-\mu_S)\Theta_\lambda(n_{n+1})\right)\times\\\left(
 \theta(\rho-p^0_{n+1}) +
\theta(p^0_{n+1}-\rho) \theta(n_{n+1}\in \text{in})-1\right)u(p_1,...,p_n) \ ,
\end{multline}
where we have generically identified the collinear cutoff prescription
with $\Theta_\lambda(n_{n+1})$ and the emission's direction with
$n_{n+1}$. We assume that the observable probes a veto on energies
less than $\rho$ in the angular out region, complementary to the in
region. If we identify the energy resolution $\mu_S$ with the
observable resolution $\rho$ then in fact the difference vanishes
except if the emission becomes collinearly unresolved in the out
region. In this case we expect that the change in between different
collinear cutoffs will need to be compensated for by using the cutoff
anomalous dimension Eq.~\ref{eq:cutoffad} to obtain a stable result.

\section{Aspects of a Hadronization Model}
\label{sec:hadronization}

We will now focus on the evolution equation for ${\mathbf S}_{n}$. In
fact it is interesting to make explicit that this object is expected
to convert the hard partons plus $n$ emissions into $m$ hadrons
eventually, and thus we write ${\mathbf S}_{n}\to {\mathbf S}_{nm}$ to
highlight this fact and assume that observables will eventually be
obtained at hadron level by integrating over the m final state
momenta.  Notice that the soft evolution is quite different in its
form as compared to the evolution equation of ${\mathbf A}_n$. In
particular, the cross-feed with different partonic multiplicities is
now coming in from larger multiplicities, with the emission operators
appearing in a form that they project down the $n+1$ onto the
$n$-parton colour space; also the anomalous dimensions are now
appearing conjugate to the case of the operator ${\mathbf A}_n$, as
otherwise they would not be able to be used as subtractions for the
virtual contributions. Any hadronization picture consistent with
amplitude level evolution must satisfy the evolution equation above;
in what follows we will now mainly focus on initial conditions and how
this perturbative evolution might tie in with existing models. To be
complete, though, let us write the solution (in four dimensions) as
\begin{multline}
  {\mathbf S}_{nm}(\mu) = {\mathbf W}^\dagger_{n}(\Lambda_{\text{NP}}|\mu)
  {\mathbf S}_{nm}^{\text{NP}}{\mathbf W}_{n}(\Lambda_{\text{NP}}|\mu) + \\
  \int_{\Lambda_{\text{NP}}}^\mu \frac{{\rm d}k}{k}\alpha_S(k)
      {\mathbf W}^\dagger_{n}(k|\mu)\int 
      \tilde{{\mathbf R}}_{n+1}^{(1,0)\dagger}
      {\mathbf S}_{n+1,m}(k) 
      \tilde{{\mathbf R}}_{n+1}^{(1,0)}[{\rm d}p_{n+1}]
      {\mathbf W}_{n}(k|\mu) \ ,
\end{multline}
where
\begin{equation}
  {\mathbf W}_{n}(k|\mu) = {\rm P}\exp\left(-\int_k^\mu \frac{{\rm
      d}q}{q}\alpha_S(q) \tilde{{\mathbf \Gamma}}_n^{(1)}\right)
\end{equation}
is the usual soft gluon evolution operator also entering the evolution
of the operator ${\mathbf A}_n$, and we have fixed an initial condition
\begin{equation}
  {\mathbf S}_{nm}(\Lambda_{\text{NP}}) = {\mathbf S}_{nm}^{\text{NP}} \ .
\end{equation}
We have also set $\mu_S=\mu_R$, thus including a running coupling
which could be continued to a meaningful infrared model.
Algorithmically, including such a model is not much different than the
algorithms we have already presented in \cite{DeAngelis:2020rvq}. In
fact, one would now generate the evolution of the operator ${\mathbf
  A}_n$, at the amplitude level, down to an infrared scale $\mu_S$ and
in the leading order case considered here the evolution would
essentially continue unaltered to even lower scales where it is
terminated by an insertion of ${\mathbf S}_{nm}^{\text{NP}}$. This
simple pattern will be altered at higher orders; it should otherwise
also not be considered an unnecessary complication since the above
factorization truly shows that one can retain {\it perturbative}
control of the soft scale $\mu_S$ (meaning that our final cross section
is independent of it to the order we have pinned the evolution down
to) and then systematically hand over to the evolution from a
high-energy 'end' of a hadronization model including effects such as
colour reconnection, which is mediated by the evolution operators just
as it is in the context of the evolution discussed in
\cite{Gieseke:2018gff}, and possibly related to further model
improvements discussed in the literature
\cite{Winter:2003tt,Gieseke:2017clv}.  We should also note that
depending on the {\it Ansatz} chosen for the non-perturbative initial
condition such an algorithm might not only be very efficient, but the
evolution needs to be inserted in order to arrive at a configuration
we can associate with a hadronic final state.

As a start we will try and use such a picture to make contact with the
cluster hadronization model
\cite{Gottschalk:1983fm,Webber:1983if,Gieseke:2003hm,Bahr:2008pv}. Let
us therefore assume that the initial conditions for ${\mathbf S}_{nm}$
will describe the transition from $n$ partons into $n-1$ primary
clusters\footnote{We can always add on more convolutions in order to
  describe the cluster fission process; for the time being we are
  interested in the perturbative evolution of ${\mathbf S}_{nm}$ in
  the context of a cluster picture.} (assuming a $q\bar{q}+ (n-2)g$
final state), and so we can take
\begin{equation}
  {\mathbf S}_{nm}^{\text{NP}}   = \delta_{m,2(n-1)} \mathbf{Cl}_n(w)
\end{equation}
where $\mathbf{Cl}_n(w)$ projects onto the valid cluster configurations
given by all colour flows which do not involve a 'singlet' gluon
contribution, but including, and explicitly allow, states where the
$q\bar{q}$ pair is colour connected and accompanied by a gluonic
system forming a singlet on its own. These are the configurations
which explicitly cannot originate from a leading-$N$, probabilistic
parton shower. We then take $\mathbf{Cl}_n(w)$ to be diagonal, so we can
simply weight each cluster configuration by a distribution of the
cluster decay products in phase space, say
\begin{equation}
  [\tau|\mathbf{Cl}_n(w)|\sigma] = \delta_{\tau\sigma} w_\sigma \delta_{{\text{cluster}(\sigma)}} \ ,
\end{equation}
where we take
\begin{equation}
  w_\sigma = \prod_{c_i,\bar{c}_j\text{ c.c. in }\sigma} w_{ij}
\end{equation}
to describe the individual cluster decay weights. The emission
operators will then map such a configuration onto one which is
contained by merging clusters, and we have
\begin{equation}
  [\tau|\tilde{\mathbf R}_{n+1}^\dagger\mathbf{Cl}_{n+1}(w)\tilde{\mathbf R}_{n+1}|\sigma] =
  N \delta_{\tau\sigma} \delta_{{\text{cluster}(\sigma)}}  \sum_{i,j\le n+1} \omega_{ij} w_{\sigma\backslash n+1}
\end{equation}
where we have abbreviated the Eikonal factor by $\omega_{ij}$, {\it i.e.}
\begin{equation}
  [\tau|\tilde{\mathbf R}_{n+1}^\dagger\mathbf{Cl}_{n+1}(w)\tilde{\mathbf R}_{n+1}|\sigma] =
  \frac{N}{2} [\tau|\mathbf{Cl}_{n}(\tilde{w}_n(w))|\sigma]
\end{equation}
where $\tilde{w}_n(w)$ links to $w$ by a simple leading-$N$ dipole
transition, which in this context we in fact associate with the
cluster fission process. The evolution operators will then mediate a
colour reconnection dynamics as discussed in \cite{Gieseke:2017clv},
however in this case the process will happen differently in the
amplitude and the conjugate amplitude. In the colour flow picture, the
`singlet' gluon (or trace condition part) would become inert to this
evolution just as it does in the hard evolution. This signals that we
need to extend the non-perturbative initial condition in fact to
include a process which describes the transition of the `singlet'
colour flow gluon into a single cluster.

\section{Summary and Outlook}
\label{ref:outlook}

In this letter we have discussed how evolution equations in colour
space, specifically those for soft gluons, originate from a general
diagrammatic recursion, infrared subtraction, and the application of a
renormalization program to re-arrange the individual contributions
into a finite cross section. We have particularly paid attention to
the role of the observable and the accuracy which the algorithm can
deliver, and to the consequences of observables which are forced to
resolve colour, either because we are considering hadronization
corrections or factorize the partonic contributions in a more general
way than needed for a jet cross section. This provided us with a new
notion of the structure of power corrections which now appear encoded
in an operator relation in colour space. They will only simplify to
more standard approaches in case of jet cross sections and by
enforcing unitarity in the sense that infrared subtractions should
relate between real and virtual contributions in a one-to-one
correspondence. We stress that, due to the nature of unstable
particles in the case of electroweak evolution \cite{Platzer:2022nfu},
this can not be achieved any longer since real and virtual corrections
are not lined up anymore, and explicit projections on isospin states
are performed by the measurement. The role of the measurement needs to
be clarified as an integral part of the evolution. We will address
this aspect in more detail in future work.

The role of the soft, or measurement operator, is to encode how
virtual and real contributions are re-arranged to cancel in the
physical cross section. For QCD, this corresponds precisely to what we
expect a hadronization model to do since in the presence of such
dynamics infrared cancellations are obscured. On a more general
ground, we can adopt the following analysis of the cross section,
which we here present in a more sketchy way with details to be
addressed later:
\begin{equation}
  \sigma = \sum_{n,m} \int\int {\rm Tr}_n \left[{\mathbf M}_n
    {\mathbf U}_{nm}\right] {\rm d}\phi_n u(\phi_m){\rm d}\phi_m
\end{equation}
is the general cross section we would strive to calculate, in terms of
$m$ experimentally observed particles from which we can calculate an
observable $u(\phi_m)$, and the trace refers to a sum over degrees of
freedom within the $n$ constituent particles from which we could build
up the $m$ observed particles. Up until now we have collectively
addressed
\begin{equation}
  {\mathbf U}_n = \sum_m \int {\mathbf U}_{nm} u(\phi_m){\rm d}\phi_m
\end{equation}
as the measurement function. The above clearly must be a consequence
of a factorization theorem which would start from general cross
section and then exploits the properties of (classes of) hadronic
observables. We therefore assume that our formalism will provide a
generalization of cross sections with many identified hadrons or the
track function formalism \cite{Chang:2013rca} to cases beyond purely
collinear physics and in particular if correlations correlators among
the hadrons are involved. Ultimately, we would start from a fully
exclusive final state with $S$ matrix element squared of the form
\begin{multline}
  \left|\langle f|S|i\rangle\right|^2 =
  \sum_{\alpha,\beta} R_{m} \int [{\rm d}p]_n
      [{\rm d}\bar{p}]_{\bar{n}} \prod_i \delta\left(\sum_{j=1}^{n_i}
      p_{ij}-P_i\right) \delta\left(\sum_{j=1}^{\bar{n}_i}
      \bar{p}_{ij}-P_i\right)\\
      \chi_{\alpha}(\{p\}_n|\{P\}_m)
      \bar{\chi}_\beta(\{\bar{p}\}_{\bar{n}}|\{P\}_m)
      \left. G^{\alpha}(p_n)\bar{G}^{\beta}(\bar{p}_{\bar{n}})\right|_{\text{$m$-(truncated,
          on-shell)}}
\end{multline}
where the sum over the field indices $\alpha,\beta$ does constitute
${\rm Tr}_n$, and the relevant Green's function is a tensor in all of
these indices, subject to choosing a basis $|\alpha\}$ of colour and
spin (which we here have deliberately denoted with a curly bracket
notation to distinguish them from the true quantum mechanical states):
$|m\rangle = |i\rangle |f\rangle$ is the product of initial and final
states involved in the definition of the $S$ matrix element, and
$\chi_{\alpha}(p_n)$ are collectively denoting the external wave
functions, {\it i.e.} in general contractions of interpolating fields
with the physical external states, or (conjugates of) Bethe-Salpeter
amplitudes for the states of interest. They are dictated by the
interpolating fields $\psi_\alpha$ we have chosen for the elementary
or possibly composite external states, and one can also consider the
optical theorem in order to arrive at a similar structure. $R_{m}$ is
the product of the corresponding wave function renormalizations for
the final state $m$, and the integrations over the off-shell
constituent momenta are constrained such that their sum equal the
observed final state momenta $P_i$. The truncation of the Green's
functions is to be understood such that iterations of
$N_m$-PI-irreducible legs which combine into a composite state of
$N_m$ constituents have been truncated. Momentum conservation of the
observed particles then results in an integration over
\begin{equation}
  \int [{\rm d}P]_m \delta\left(\sum_{i=1}^m P_i - Q\right)\prod_{i=1}^m \tilde{\delta}\left(P_i^2,M_i^2\right) \ .
\end{equation}
Our current formalism has then assumed that the result will be
dominated by those contributions which have all partonic lines put
on-shell,
\begin{equation}
  \{\alpha| {\mathbf M}_n |\beta \} = (R_{n,\alpha}R_{\bar{n},\beta})^{1/2}
  \left. G^{\alpha}(p_n)\bar{G}^{\beta}(\bar{p}_{\bar{n}})\right|_{\text{1-(truncated, on-shell)}} \ ,
\end{equation}
and
\begin{equation}
  \{\beta | {\mathbf U}_{nm} |\alpha\} =
  \frac{R_{m}}{(R_{n,\alpha}R_{\bar{n},\beta})^{1/2}}
      \chi_{\alpha}(\{p\}_n|\{P\}_m)
      \bar{\chi}_\beta(\{\bar{p}\}_{\bar{n}}|\{P\}_m) u(\{P\}_m) \ ,
\end{equation}
where each of the $P_i$ is given by the sum of constituents $\sum_j
p_{ij}$ and the two factors are then convoluted by momentum integrations
\begin{equation}
{\cal R} =   \prod_{i=1}^m \delta\left(\sum_{j=1}^{n_i} p_{ij} - \sum_{j=1}^{\bar{n}_i} \bar{p}_{ij}\right)
  \tilde{\delta}\left(\sum_{j=1}^{n_i} p_{ij},M_i^2\right)\prod_{j=1}^{\bar{n}_i} \tilde{\delta}(\bar{p}_{ij},\bar{m}_{ij}^2)
        [{\rm d}\bar{p}_{ij}] \ ,
\end{equation}
whereas the remaining integration then is over the partonic phase space
\begin{equation}
  \int [{\rm d}p]_m \delta\left(\sum_{i=1}^m\sum_{j=1}^{n_i}
  p_{ij}- Q\right)\prod_{i=1}^m\prod_{j=1}^{n_i} \tilde{\delta}\left(p_{ij}^2,m_{ij}^2\right) \ .
\end{equation}
Subject to the observable and leading regions the convolution measure
can then be simplified further to yield the soft gluon algorithm we
have been discussing now. It is worth noting that ${\cal R}$ leaves
$(d-1)\left(\sum_{i=1}^m \bar{n}_i - m\right)$ degrees of freedom to
integrate over. Essentially the soft gluon algorithm can be
constructed by transforming, for $n_i=\bar{n}_i$, the $\bar{p}_{ij}$
into scaled versions close to the $p_{ij}$ and then to expand around
scaling parameters near one (see
\cite{Loschner:2021keu,Platzer:2022nfu} for more discussion on such
mappings) which then do not appear in the conjugate amplitude anymore
and can easily be integrated over. While we have been making no
assumptions about the precise form of ${\mathbf U}$ and have studied
its mere presence, a careful analysis of the above then clearly
reveals an operator definition. Also note that in this context, the
soft factor can comprise operators with more particles overlapping
with the physical, external states -- this exactly corresponds to the
ability of ${\mathbf S}_n$ to absorb emissions which have been
generated during the evolution of the hard process as the emission
contribution in the evolution of ${\mathbf S}_n$ corresponds to a
transition from a larger to a smaller final state.

Beyond the soft gluon case we would need to allow the algorithm to
populate momentum configurations and flavours possibly differently in
between the amplitude and its conjugate so long as they serve to
provide constituents consistent with the measured final states (in the
sense of interpolating fields with non-vanishing overlaps). This might
already apply in the case of hard collinear physics, where
convolutions not only exist in the longitudinal momentum fractions but
also regarding transverse momenta. The expansion around the partonic
mass shells is then still possible, and a treatment of the algorithms
along these lines will become crucial in the electroweak case. It will
heavily rely on the fact that we maintain overall energy-momentum
conservation in the amplitude and the conjugate, as advocated in
\cite{Platzer:2022nfu}. The main purpose of this analysis and outlook
is, however, to show that we can touch ground with first-principle
definitions, an avenue which we will explore further. Beyond this
observation we note that correlations among hadrons with certain
momentum balances might actually be able to probe whatever (colour)
interference the hard process allows to, since this will be imprinted
in the evolution much in the same way as it has been observed in the
case of double parton distributions for those processes where colour
interference will mix these objects \cite{Diehl:2021wvd}. In fact, the
general formula we envisage should be easily able to accommodate
processes with incoming hadrons.

\section*{Acknowledgments}

I am deeply grateful to Ines Ruffa and Maximilian L\"oschner to
uncountable comments and discussions. I am also grateful to Reinhard
Alkofer, Thomas Becher, Jeff Forshaw, Jack Holguin, Massimiliano
Procura and Malin Sj\"odahl for valuable discussions. This work has
been supported in part by the European Union’s Horizon 2020 research
and innovation programme as part of the Marie Skłodowska-Curie
Innovative Training Network MCnetITN3 (grant agreement no. 722104),
and in part by the COST actions CA16201 ``PARTICLEFACE'' and CA16108
``VBSCAN''. I am grateful to the MITP programmes at Mainz where this
work has been initiated, and to the Erwin Schr\"odinger Institute
Vienna for hospitality and support while this work has been finalized
within the Research in in Teams Programmes ``Higher-order Corrections
to Parton Branching at the Amplitude Level'' (RIT2020), ``Amplitude
Level Evolution I: Initial State Evolution'' (RIT0421), and
``Amplitude Level Evolution II: Cracking down on colour bases.''
(RIT0521).

\appendix

\section{Factorization and removal of over-subtractions}
\label{sec:oversubtract}

Identifying subtraction terms at the leading order is rather
straightforward, however at the next order there will be two
contributions which mix different partonic multiplicities or different
loop orders. In the limit that the maximum number of emissions or loop
momenta ranges into the singular regions, an over-subtraction is thus
taking place which needs to be removed. This is not unique, however in
the present work we choose to arrange things such that the remaining
contribution will be finite as dictated by the resolution function.
Expanding up to the first order we only need to consider $\hat{\mathbf
  M}_{n}^{(0)}$ and $\hat{\mathbf M}_{n}^{(1)}$, and in this case we
have
\begin{eqnarray}
  {\cal S}^{(1)}_{\text{tree}}\left[{\mathbf M}_{n,R}^{(0)}\right] &=&
  {\mathbf F}_n^{(1,0)} {\mathbf M}^{(0)}_{n-1,R} {\mathbf F}_n^{(1,0)\dagger} \ ,\\\nonumber
  {\cal S}^{(0)}_{\text{1-loop}}\left[{\mathbf M}_{n,R}^{(1)}\right] &=&
  {\mathbf X}_n^{(1)} {\mathbf M}^{(0)}_{n,R}
  + {\mathbf M}^{(0)}_{n,R} {\mathbf X}_n^{(1)\dagger} \ .
\end{eqnarray}
Thus, observing that in the singly unresolved limit ({\it i.e.} here a
single soft emission or exchange at leading power)
\begin{eqnarray}
  \label{eq:singlyunresolved}
  {\mathbf M}_{n,R}^{(0)} &\to& \hat{\mathbf D}^{(1,0)}_n {\mathbf M}_{n-1,R}^{(0)} \hat{\mathbf D}^{(1,0)\dagger}_n\\\nonumber
  {\mathbf M}_{n,R}^{(1)} &\to& \hat{\mathbf V}_n^{(1)}{\mathbf M}_{n,R}^{(0)} + {\mathbf M}_{n,R}^{(0)} \hat{\mathbf V}_n^{(1)\dagger}
  + \hat{\mathbf D}^{(1,0)}_n {\mathbf M}_{n-1,R}^{(1)} \hat{\mathbf D}^{(1,0)\dagger}_n \ ,
\end{eqnarray}
which {\it defines} $\hat{\mathbf D}^{(1,0}_n$ and $\hat{\mathbf
  V}^{(1)}_n$, we can take ${\mathbf F}_n^{(1,0)}\circ {\mathbf
  F}_n^{(1,0)\dagger}$ and ${\mathbf X}_n^{(1)}$ as in
Sec.~\ref{sec:infrared} to obtain finite matrix elements
\begin{eqnarray}
 \hat{\mathbf M}_{n}^{(0)} &=&  {\mathbf M}_{n,R}^{(0)} - {\cal S}^{(1)}_{\text{tree}}\left[{\mathbf
      M}_{n,R}^{(0)}\right] \\\nonumber &\to&
   (1-\Theta_{n,1})
   \hat{\mathbf D}^{(1,0)}_n {\mathbf
     M}_{n-1,R}^{(0)} \hat{\mathbf D}^{(1,0)\dagger}_n\\\nonumber
    \hat{\mathbf M}_{n}^{(1)} &=&  {\mathbf M}_{n,R}^{(1)} - {\cal S}^{(1)}_{\text{tree}}\left[{\mathbf
      M}_{n,R}^{(1)}\right]- {\cal S}^{(1)}_{1\text{-loop}}\left[{\mathbf
        M}_{n,R}^{(0)}\right]\\\nonumber
    &\to& 
   (1-\Theta_{n,1})
   \hat{\mathbf D}^{(1,0)}_n {\mathbf
     M}_{n-1,R}^{(1)} \hat{\mathbf D}^{(1,0)\dagger}_n
   \\\nonumber & + &  (1-\xi \Theta_{n,1})\left(\hat{\mathbf
       V}_n^{(1)}\left[1-\Xi_{n,1}\right]{\mathbf M}_{n,R}^{(0)} +
           {\mathbf M}_{n,R}^{(0)} \hat{\mathbf
             V}_n^{(1)\dagger}\left[1-\Xi_{n,1}\right] \right) \ ,
\end{eqnarray}
which now regulates both the virtual contributions as well as the real
contributions in a singly unresolved limit of the $n$'th emission or
the last ({\it i.e.} outermost) virtual exchange operator. Notice if
$\xi=0$ then the contribution from the factored virtuals will not be
finite; interestingly, this subtraction will still lead to finite
anomalous dimension.  Since the renormalized matrix elements are free
of UV divergences our subtractions shall only cover infrared singular
regions. The definition of the renormalized factorized virtual
corrections can be obtained using subtractions which facilitate UV
counterterms as given in \cite{Becker:2010ng,Capatti:2022tit}. The
subtractions, however, also need to remove infinite momenta of the
soft gluon, which originate purely from the approximation we use.  At
the second order, over-subtraction can occur in multiply unresolved
limits. As an example for the removal of over-subtractions consider
the real emission subtractions up to second order. In this case the
subtraction at fixed multiplicity is
\begin{equation}
  {\cal S}^{(2)}_{\text{tree}}\left[{\mathbf M}_{n,R}^{(0)}\right] =
  {\mathbf F}^{(1,0)} {\mathbf M}^{(0)}_{n-1,R} {\mathbf F}^{(1,0)\dagger}
  + {\mathbf F}^{(2,0)} {\mathbf M}^{(0)}_{n-2,R} {\mathbf F}^{(2,0)\dagger} \ .
\end{equation}
At the first order we would have identified ${\mathbf F}^{(1,0)} \circ
{\mathbf F}^{(1,0)\dagger} = \hat{\mathbf D}^{(1,0)} \circ
\hat{\mathbf D}^{(1,0)\dagger}\Theta_{n,1}$. In the doubly-unresolved
limit (emission of two soft gluons, exchange of two soft gluons, or
emission and exchange of one soft gluon) we then need to account for
${\mathbf M}_{n,R}$ factoring onto ${\mathbf M}_{n-2,R}$,
\begin{equation}
  {\mathbf M}_{n,R}^{(0)} \to \hat{\mathbf D}^{(2,0)}_n {\mathbf M}_{n-2,R}^{(0)} \hat{\mathbf D}^{(2,0)\dagger}_n
\end{equation}
and using the single-emission factorization, to yield
\begin{multline}
  {\mathbf M}_{n,R}^{(0)} - {\cal S}_{\text{tree}}^{(2)}\left[{\mathbf M}_{n,R}^{(0)}\right] \to
  \hat{\mathbf D}_n^{(2,0)}{\mathbf M}^{(0)}_{n-2,R}\hat{\mathbf D}_n^{(2,0)\dagger}\\
  - \hat{\mathbf D}_n^{(1,0)}\hat{\mathbf D}_{n-1}^{(1,0)} {\mathbf M}^{(0)}_{n-2,R} \hat{\mathbf D}_{n-1}^{(1,0)\dagger}\hat{\mathbf D}_n^{(1,0)\dagger}\Theta_{n,1}
  - {\mathbf F}^{(2,0)} {\mathbf M}^{(0)}_{n-2,R} {\mathbf F}^{(2,0)\dagger} \ .
\end{multline}
Demanding that
\begin{equation}
  {\mathbf M}_{n,R}^{(0)} - {\cal S}^{(2)}_{\text{tree}}\left[{\mathbf M}_{n,R}^{(0)}\right] \to
  \hat{\mathbf D}_n^{(2,0)}{\mathbf M}^{(0)}_{n-2,R}\hat{\mathbf D}_n^{(2,0)\dagger}(1-\Theta_{n,2})
\end{equation}
then fixes the second-order subtraction to what we have been
advocating in the main text. Similar remarks apply to the other
subtractions, in particular we have for the second order one-loop
subtraction
\begin{multline}
  {\mathbf M}_{n,R}^{(1)}- {\cal S}^{(2)}_{\text{tree}}\left[{\mathbf
      M}_{n,R}^{(1)}\right]- {\cal
    S}^{(2)}_{1\text{-loop}}\left[{\mathbf M}_{n,R}^{(0)}\right] \to
   \hat{\mathbf D}_n^{(2,0)}{\mathbf M}^{(1)}_{n-2,R}\hat{\mathbf
    D}_n^{(2,0)\dagger}(1-\Theta_{n,2})\\+
  \left(\hat{\mathbf D}_n^{(1,1)}\left[1-\Xi_{n,1}\right] {\mathbf
    M}_{n-1,R}^{(0)}{\mathbf D}_n^{(1,0)\dagger} + \hat{\mathbf
      D}_n^{(1,0)} {\mathbf M}_{n-1,R}^{(0)}\hat{\mathbf
        D}_n^{(1,1)\dagger}\left[1-\Xi_{n,1}\right]\right)(1-\Theta_{n,1})
\end{multline}
where we have used that
\begin{equation}
  {\mathbf M}_{n,R}^{(1)} \to \hat{\mathbf D}_{n}^{(2,0)}{\mathbf M}_{n-2,R}^{(1)}\hat{\mathbf D}_{n}^{(2,0)\dagger}
  + \hat{\mathbf D}_{n}^{(1,1)}{\mathbf M}_{n-1,R}^{(0)}\hat{\mathbf D}_{n}^{(1,0)\dagger}
    + \hat{\mathbf D}_{n}^{(1,0)}{\mathbf M}_{n-1,R}^{(0)}\hat{\mathbf D}_{n}^{(1,1)\dagger}
\end{equation}
and Eq.~\ref{eq:singlyunresolved} in order to compare all quantities
to the same (leading) power. As we are now in possession of the tree
level and one-loop subtraction terms, we can finally determine the
two-loop contributions. Here we use
\begin{eqnarray}
  {\mathbf M}_{n,R}^{(2)} &\to& \hat{\mathbf D}_{n}^{(2,0)}\hat{\mathbf M}_{n-2,R}^{(2)}\hat{\mathbf D}_{n}^{(2,0)\dagger}\\\nonumber
  &+& \hat{\mathbf D}_{n}^{(1,1)}\hat{\mathbf M}_{n-1,R}^{(1)}\hat{\mathbf D}_{n}^{(1,0)\dagger}
    + \hat{\mathbf D}_{n}^{(1,0)}\hat{\mathbf M}_{n-1,R}^{(1)}\hat{\mathbf D}_{n}^{(1,1)\dagger}\\\nonumber&+&
  \hat{\mathbf V}_n^{(2)}{\mathbf M}_{n,R}^{(0)} + {\mathbf M}_{n,R}^{(0)}\hat{\mathbf V}_n^{(2)\dagger} +
  \hat{\mathbf V}_n^{(1)}{\mathbf M}_{n,R}^{(0)}\hat{\mathbf V}_n^{(1)\dagger}
\end{eqnarray}
and the other factorisation formulae. Then we find
\begin{eqnarray}
\hat{\mathbf M}_n^{(2)} &=&   {\mathbf M}_{n,R}^{(2)}- {\cal S}^{(2)}_{\text{tree}}\left[{\mathbf
      M}_{n,R}^{(2)}\right]- {\cal
      S}^{(2)}_{1\text{-loop}}\left[{\mathbf M}_{n,R}^{(1)}\right]
    - {\cal
    S}^{(2)}_{2\text{-loop}}\left[{\mathbf M}_{n,R}^{(0)}\right] \\\nonumber &\to &
   \hat{\mathbf D}_n^{(2,0)}{\mathbf M}^{(2)}_{n-2,R}\hat{\mathbf
    D}_n^{(2,0)\dagger}(1-\Theta_{n,2})\\\nonumber &+&
  \left(\hat{\mathbf D}_n^{(1,1)}\left[1-\Xi_{n,1}\right] {\mathbf
    M}_{n-1,R}^{(1)}\hat{\mathbf D}_n^{(1,0)\dagger} + \hat{\mathbf
      D}_n^{(1,0)} {\mathbf M}_{n-1,R}^{(1)}\hat{\mathbf
    D}_n^{(1,1)\dagger}\left[1-\Xi_{n,1}\right]\right)(1-\Theta_{n,1})\\\nonumber 
  &+& \left(1-\xi\Theta_{n,1}\right)^2\hat{\mathbf V}^{(1)}_n \left[1-\Xi_{n,1}\right] {\mathbf M}_{n,R}^{(0)}
  \hat{\mathbf V}^{(1)\dagger}_n \left[1-\Xi_{n,1}\right] \\\nonumber
 &+& (1-\xi \Theta_{n,2})\left(\hat{\mathbf V}^{(2)}_n \left[1-\Xi_{n,2}\right]{\mathbf M}_{n,R}^{(0)}(1-\xi \Theta_{n,2})+
  {\mathbf M}_{n,R}^{(0)}\hat{\mathbf V}^{(2)\dagger}_n \left[1-\Xi_{n,2}\right]\right) \ ,
\end{eqnarray}
which is again totally finite, in particular also for $\xi=1$.

\section{Complete Results for Anomalous Dimensions and Emission Operators}

\subsection{Hard Density Operator}
\label{sec:anomhard}

If we write
\begin{equation}
  {\mathbf \Gamma}_{S,n} = \sum_{l\ge 1} \alpha_S^l {\mathbf \Gamma}_{S,n}^{(l)}\qquad
  {\mathbf R}_{S,n}^{(s)} \circ {\mathbf R}_{S,n}^{(s)\dagger} = \sum_{l\ge 1}\sum_{l'=0}^{l-1}\alpha_S^l
  {\mathbf R}_{S,n}^{(s,l-l')} \circ {\mathbf R}_{S,n}^{(s,l')\dagger}
\end{equation}
we have
\begin{equation}
  {\mathbf \Gamma}_{S,n}^{(1)} = \frac{n}{2}\beta_{S,0}  - \beta_{S,-1}{\mathbf Z}_n^{(1)}  - \partial_S {\mathbf Z}_n^{(1)}
\end{equation}
  \begin{equation}
    {\mathbf \Gamma}_{S,n}^{(2)} = \frac{n}{2} \beta_{S,1}+\beta_{S,-1}\left(
    -2 {\mathbf Z}_n^{(2)} + \left({\mathbf Z}_n^{(1)}\right)^2\right)
    -{\mathbf Z}_n^{(1)}\beta_{S,0} - \left(\partial_S{\mathbf Z}_n^{(2)} - {\mathbf Z}_{n}^{(1)}\partial_S {\mathbf Z}_{n}^{(1)}\right)
\end{equation}
  as well as
  \begin{eqnarray}
    {\mathbf R}_{S,n}^{(1,0)} \circ {\mathbf R}_{S,n}^{(1,0)\dagger} &
    = & \partial_S\left({\mathbf E}_{n}^{(1,0)}\circ {\mathbf
      E}_{n}^{(1,0)\dagger}\right) + \beta_{S,-1}{\mathbf
      E}_{n}^{(1,0)}\circ {\mathbf E}_{n}^{(1,0)\dagger} \\ {\mathbf
      R}_{S,n}^{(1,1)} \circ {\mathbf R}_{S,n}^{(1,0)\dagger} & = &
    \partial_S \left({\mathbf E}_n^{(1,1)}\circ {\mathbf
      E}_n^{(1,0)\dagger}\right) - {\mathbf Z}_n^{(1)}\partial_S\left(
            {\mathbf E}_n^{(1,0)}\circ {\mathbf
              E}_n^{(1,0)\dagger}\right) \\\nonumber &+& {\mathbf
              E}_n^{(1,0)}{\mathbf \Gamma}_{S,n-1}^{(1)}\circ {\mathbf
              E}_n^{(1,0)\dagger} + \beta_{S,-1}\left({\mathbf
              E}_n^{(1,1)} -\left( {\mathbf Z}_n^{(1)} + \frac{n+1}{2}
            Z_g^{(1)}\right) {\mathbf E}_n^{(1,0)} \right)\circ
            {\mathbf E}_n^{(1,0)\dagger}
  \end{eqnarray}
  and
  \begin{equation}
    {\mathbf R}_{S,n}^{(2,0)} \circ {\mathbf R}_{S,n}^{(2,0)\dagger} =
    \partial_S\left( {\mathbf E}_{n}^{(2,0)} \circ {\mathbf E}_{n}^{(2,0)\dagger}\right) -
            {\mathbf E}_{n}^{(1,0)} {\mathbf R}_{S,n-1}^{(1,0)} \circ {\mathbf R}_{S,n-1}^{(1,0)\dagger}{\mathbf E}_{n}^{(1,0)\dagger}
            + 2\beta_{S,-1} {\mathbf E}_{n}^{(2,0)} \circ {\mathbf E}_{n}^{(2,0)\dagger}
  \end{equation}

  \subsection{Soft Function}
  \label{sec:anomsoft}
  
Up to second order we find in this case:
\begin{eqnarray}
  \tilde{\mathbf \Gamma}_{S,n}^{(1)} &=& -\beta_{S,-1}{\mathbf X}_n^{(1)} - \partial_S {\mathbf X}_n^{(1)} \\
  \tilde{\mathbf \Gamma}_{S,n}^{(2)} &=&  -\beta_{S,-1}\left( 2{\mathbf X}_n^{(2)}+ ({\mathbf X}_n^{(1)})^2  \right)- \beta_{S,0} {\mathbf X}_n^{(1)} -
  \partial_S{\mathbf X}_n^{(2)} -
        \left(\partial_S{\mathbf X}_n^{(1)}\right){\mathbf X}_n^{(1)}
\end{eqnarray}
\begin{eqnarray}
  \mu_R^{-2\epsilon}\tilde{\mathbf R}_{S,n}^{(1,0)\dagger}\circ \tilde{\mathbf R}_{S,n}^{(1,0)} &=&
  \partial_S\left({\mathbf F}_{n}^{(1,0)\dagger}\circ{\mathbf F}_{n}^{(1,0)}\right) \\
  \mu_R^{2\epsilon}\tilde{\mathbf R}_{S,n}^{(1,1)\dagger}\circ \tilde{\mathbf R}_{S,n}^{(1,0)} &=&
  \partial_S \left(\tilde{\mathbf F}_{n}^{(1,1)\dagger}\circ \tilde{\mathbf F}_{n}^{(1,0)}\right)
  + {\mathbf X}_n^{(1)\dagger}\partial_S \left(\tilde{\mathbf F}_{n}^{(1,0)\dagger}\circ \tilde{\mathbf F}_{n}^{(1,0)}\right)\\\nonumber
  &-& \tilde{\mathbf F}_{n}^{(1,0)\dagger}\tilde{\mathbf \Gamma}_{S,n}^{(1)}\circ \tilde{\mathbf F}_{n}^{(1,0)}
  +\beta_{S,-1}\left(\tilde{\mathbf F}_n^{(1,1)} +\frac{1}{2} Z_g^{(1)}
                             \tilde{\mathbf F}_n^{(1,0)}
                             \right)\circ \tilde{\mathbf F}_n^{(1,0)\dagger}
\end{eqnarray}
\begin{eqnarray}
  \mu_R^{-4\epsilon}\tilde{\mathbf R}_{S,n}^{(2,0)\dagger}\circ \tilde{\mathbf R}_{S,n}^{(2,0)} &=&
  \partial_S\left({\mathbf F}_{n}^{(2,0)\dagger}\circ{\mathbf F}_{n}^{(2,0)}\right)
  + {\mathbf F}_{n-1}^{(1,0)\dagger} \tilde{\mathbf R}_{S,n}^{(1,0)\dagger}\circ \tilde{\mathbf R}_{S,n}^{(1,0)}
  {\mathbf F}_{n-1}^{(1,0)}\mu_R^{-2\epsilon}
\end{eqnarray}

\bibliography{hadronization}

\end{document}